\documentclass[aps,pre,twocolumn,groupedaddress,showpacs,superscriptaddress,amssymb,amsmath]{revtex4-2}
\usepackage{graphicx}
\usepackage{dcolumn}
\usepackage{bm}
\usepackage{hyperref}
\usepackage{cleveref}
\hypersetup{
    colorlinks=true,
    linkcolor=blue,
    urlcolor=blue,
	citecolor=blue
}
\usepackage{comment}
\usepackage{color}
\usepackage[utf8]{inputenc}
\usepackage{graphicx}
\usepackage{tabularx}
\usepackage{xcolor}
\usepackage{amsmath}
\usepackage{dcolumn}
\usepackage{hyperref}
\usepackage{bm}
\usepackage{epsf}
\usepackage{braket}
\usepackage{tensor}
\usepackage{soul}

\begin{document}
\newcolumntype{M}[1]{>{\centering\arraybackslash}m{#1}}

\title{Transport in open quantum systems in presence of lossy channels}

\author{Katha Ganguly}
\email{katha.ganguly@students.iiserpune.ac.in}
\affiliation{Department of Physics, Indian Institute of Science Education and Research Pune, Dr. Homi Bhabha Road, Ward No. 8, NCL Colony, Pashan, Pune, Maharashtra 411008, India}

\author{Manas Kulkarni}
\email{manas.kulkarni@icts.res.in} 
\affiliation{International Centre for Theoretical Sciences, Tata Institute of Fundamental Research,
Bangalore 560089, India}

\author{Bijay Kumar Agarwalla}
\email{bijay@iiserpune.ac.in}
\affiliation{Department of Physics, Indian Institute of Science Education and Research Pune, Dr. Homi Bhabha Road, Ward No. 8, NCL Colony, Pashan, Pune, Maharashtra 411008, India}

\date{\today}

\begin{abstract}
We study nonequilibrium steady state (NESS) transport in a boundary driven one-dimensional fermionic lattice setup which is further subjected to particle loss. We analyze the system size scaling of conductance at zero temperature for different values of the chemical potential of the boundary reservoirs. We consider a variety of loss channel configurations: (i) single loss at the middle site of the lattice, (ii) multiple but nonextensive lossy channels, and (iii) extensive lossy channels. For the cases (i) and (ii), the conductance scaling with system size remains robust (i.e., same as the case with no loss) for chemical potential within and outside the lattice band, while at the band-edge rich anomalous conductance scaling emerges. For case (iii), the conductance scaling becomes ballistic in the thermodynamic limit for any value of chemical potential. We explain the emergence of these different system size scalings of conductance by analyzing the spectral properties of the associated non-hermitian transfer matrices of the underlying lattice. We demonstrate that the emergence of anomalous scaling is deeply connected to the existence of exceptional points of transfer matrices. Our study unravels that by carefully optimizing the loss mechanism configurations, one can in principle realize systems with rich transport properties in low-dimensional open quantum systems.

\end{abstract}

\maketitle

\section{Introduction}
\label{sec:Introduction}
Transport in low-dimensional open quantum systems with boundary drives has been an active area of research both theoretically \cite{nazarov2009quantum, Landi-review-2022, wang_review, Dubi_review,Abhishek_heat_review,LIVI_review,Heat_transport_Xu,Znidaric_review,Purkayastha_transport,Purkayastha_2019,PhysRevA.105.032208,PhysRevA.107.062216,annurev_dvira,Galperin_2007,madhumita_anomalous,YANG201085,segal_transport,Benenti_review} and experimentally \cite{Latha_transport,Ivana_transport_exp,Dadosh_nature,Atala_transport,Shuan_electron_transport,electron_transport_Nitzman,Kouwenhoven1997,doi:10.1126/science.aac9584,molecule_transport,ZHANG20201,Thermal_expt_Gotsmann}. The system which is connected to boundary drives can potentially have additional dissipative channels which arise due to inevitable imperfections in experimental platforms. These dissipation channels are not necessarily detrimental. In fact, rapid progress in quantum technologies \cite{MULLER20121,2022NatRP...4..660H,article,PhysRevLett.116.235302,Ott_2016,doi:10.1126/sciadv.1701513,Blatt_hydro} has made it possible to engineer such channels to one's advantage and manipulate the system to obtain the desired functionalities \cite{Damanet_2019,Zaikin_transport,PhysRevA.89.013620}. 

One of the most common types of dissipation present in a typical electron transport setup is due to electron-phonon interactions \cite{PhysRevLett.113.236603,chen_ep}. This is often modeled via a dephasing mechanism where the channel results in energy loss/gain without any loss/gain of particles \cite{PhysRevLett.46.211,PhysRevB.75.195110,PhysRevLett.122.050501,PhysRevLett.123.180402,PhysRevB.97.214202}. Presence of such dephasing channel throughout the setup often leads to remarkable change in transport properties \cite{Znidaric2013, PhysRevB.93.094205, PhysRevB.80.125423, PhysRevB.82.144201, PhysRevB.102.100301, PhysRevB.80.125423, PhysRevB.82.144201, nature_michael, PhysRevResearch.2.023294, PhysRevResearch.3.013086, PhysRevB.100.155431, Ghosh_2024}. A particularly well studied aspect is that of how transport scales with system size \cite{PhysRevB.104.174203,PhysRevB.109.165408}. These dephasing channels are not only known
to render the system diffusive by mimicking scattering mechanism, but also can result in anomalous transport by deviating from the conventional Ohm's law \cite{PhysRevB.109.165408,dhawan2024anomaloustransportlongrangedopen}. Moreover, presence of such dephasing mechanism can also lead to enhancement of transport when the setup is carefully designed, thereby turning imperfection to an advantage \cite{PhysRevA.104.022205,PhysRevB.104.144301,superballistic_madhumita}.

Another important kind of dissipative mechanism is the actual loss of particles from the system. This is often unavoidable but can also be carefully engineered to one's advantage. For bosonic systems, such particle loss has been realized in several experiments by applying electron beams on atomic Bose-Einstein condensates \cite{PhysRevLett.110.035302,PhysRevLett.102.144101} and also for interacting bosonic gases ~\cite{PhysRevLett.116.235302}. Moreover, recently in cold atom platforms, localized loss of fermions has been realized using optical tweezers \cite{PhysRevA.100.053605,PhysRevLett.123.193605,PhysRevLett.130.200404}. 
In parallel to these experiments, there have been intense theoretical progress in understanding such lossy systems \cite{PhysRevA.85.063620,PhysRevA.99.031601,PhysRevA.97.053614,PhysRevLett.125.240404,PhysRevResearch.3.L012016,PhysRevB.101.075139,PhysRevB.101.144301,PhysRevB.105.054303,PhysRevB.104.155431,PhysRevResearch.6.L012039,PhysRevResearch.5.033095,marco_schiro_loss}. 
Very recently, the impact of local loss on boundary driven setup was theoretically investigated \cite{Giamarchi_loss,Giamarchi_nonlinear}. The intrinsic interplay between boundary drive, local loss in the bulk, and inherent symmetry of the system was unraveled. Needless to mention, despite these recent theoretical and experimental developments, our understanding of loss-gain mechanism in quantum dynamics and quantum transport has not reached the same level of maturity in comparison to that of dephasing mechanism. One such important gap for the case of loss/gain channels is that of characterization of transport via studying system scaling of conductance which is well explored for dephasing channels \cite{PhysRevB.109.165408,PhysRevB.75.195110,madhumita_buttiker,PhysRevB.41.7411,dhawan2024anomaloustransportlongrangedopen}. 

In this work, we provide an in-depth study of conductance for a boundary driven lattice setup where the system is subjected to local, multiple, or extensive loss in the bulk. In particular, we focus on system size scaling of conductance \cite{subdiffusive_madhumita,PhysRevB.109.165408,dhawan2024anomaloustransportlongrangedopen,PhysRevB.86.125118,madhumita_buttiker,PhysRevB.41.7411} which is a hallmark for classifying different transport regimes (such as ballistic, sub/super diffusive, diffusive, localized) in open quantum systems. We observe interesting transport behaviour with system size in presence of loss when the chemical potential of the boundary reservoir is tuned. We provide extensive numerical results and compelling analytical proofs that relies on the powerful tool of the transfer matrices \cite{subdiffusive_madhumita,hypersurface_madhumita,PhysRevLett.127.240601,saha2024effectordertransfermatrix,isolated_draft,Vatsal_transfer1,Vatsal_transfer2}.

Below, we summarize our main findings:
\begin{enumerate}
\item For chemical potential values that correspond to either within the band or outside the band of the lattice, the system size ($N$) scaling of conductance remains robust  in presence of single localized loss or multiple but non-extensive number $\mathcal{O}(1)$ of lossy channels. In other words, the system size scaling of conductance remains ballistic within the band and exponentially localized with system size, outside the band, i.e., similar to what we see in the case with no lossy channels.

\item Remarkably, the system size scaling of conductance when the chemical potential is located at the band-edge is very rich in presence of both single or multiple non-extensive $\mathcal{O}(1)$ lossy channels. We observe crossover between different interesting transport regimes. The window of these regimes and the crossover length scales can be estimated and controlled by the loss strength. 

\item Furthermore, the conductance scaling in presence of extensive number $\mathcal{O}(N)$ of lossy channels does not remain robust even within and outside the band and scales ballistically in the large $N$ limit. For chemical potential at band-edge, conductance scaling shows a crossover from superballistic to ballistic regime as we increase $N$.
\end{enumerate}

We organize the paper as follows. In Sec.~\ref{sec:2}, we discuss the lattice setup which is connected to boundary reservoirs. In Sec.~\ref{sec-III}, we  discuss the case of localized loss, calculate the steady state conductance and present the results for conductance scaling with system size. We focus on different range of lattice hopping cases to unravel interesting transport properties. In Sec.~\ref{sec-IV}, we discuss the scaling of conductance in case of multiple non-extensive number of lossy channels. In Sec.~\ref{sec-5}, we extend our study by considering extensive number of lossy channels. Finally in Sec.~\ref{sec:conc}, we summarize our main results along with an outlook. Certain details are delegated to the appendices. 
\section{Setup}
\label{sec:2}
\begin{figure}
    \centering
    \includegraphics[width=8.7cm]{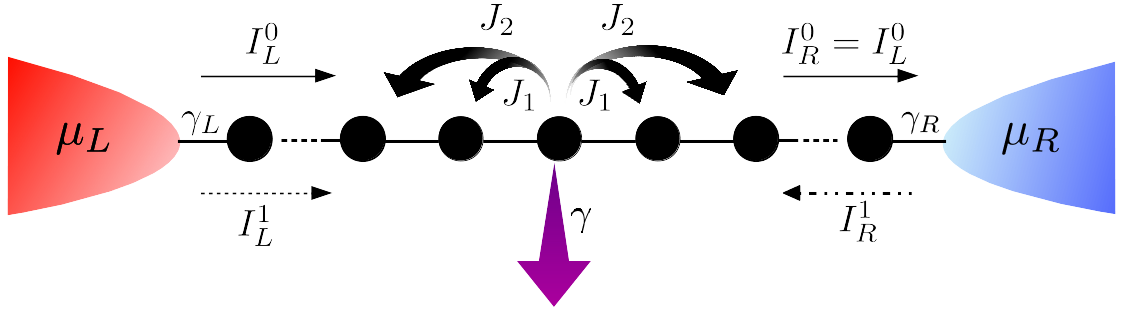}
    \caption{Schematic of the setup: one-dimensional fermionic lattice connected to left and right reservoirs at the two boundaries of the system [See Eq.~\eqref{Ham}]. The left and right reservoirs are maintained at chemical potentials $\mu_L$ and $\mu_R$, respectively, with $\mu_L>\mu_R$ and are kept at zero temperature. We consider the lattice with finite range of hopping $n$. Here in the figure we take $n=2$ with $J_1$ and $J_2$ being the nearest neighbour and next nearest neighbour hopping amplitudes. The lattice is further subjected to local particle loss from the central site of the lattice with loss rate $\gamma$ [See Eq.~\eqref{density-loss}].  $I_{L}^{0}$ is the current coming out from the left reservoir (solid arrow) and going into the right reservoir, $I_{L}^{1}$ is the current coming out from the left reservoir and lost through the dissipative channel at the middle site (Dashed arrow). $I_{R}^{0}$ which is also equal to $I_L^{0}$ is the current coming into the right reservoir from the left reservoir (solid arrow) and $I_{R}^{1}$ is the current drawn from the  right reservoir that flows through the dissipated site (Dashed-dotted arrow).}
    \label{schematic-local-loss}
\end{figure}

Let us consider a non-interacting one-dimensional fermionic lattice with finite range hopping upto $n$-th nearest neighbour. The lattice is further connected to two fermionic reservoirs one at each boundary (left and right) which are kept at a chemical potential $\mu_L$ and $\mu_R$, respectively (see schematic Fig.~\ref{schematic-local-loss}). The Hamiltonian for this setup is given by
\begin{align}
H=H_{S}+H_{L}+H_{R}+H_{LS}+H_{RS}, \label{Ham}
\end{align}
where $H_{S}$ is the Hamiltonian of the system i.e. a one dimensional fermionic lattice, given by (setting $\hbar =1$)
\begin{align}
    H_{S}=-\sum_{i=1}^{N}\sum_{r=1}^{n} J_{r}\big(c^{\dagger}_{i}c_{i+r}+h.c.\big), \label{finite-range-Ham}
\end{align}
where $J_r$ is the hopping amplitude from $i$-th site to $i+r$-th site, $n$ is the range of hopping, and $N$ is the lattice size which is taken to be odd here.  $c_i \,(c_i^{\dagger})$ is the fermionic annihilation (creation) operator at the $i$-th site. In the thermodynamic limit $(N \to \infty)$ the single-particle dispersion relation for the lattice system is given by
\begin{align}
\label{dispersion}
    \omega(k)=-2\sum_{r=1}^{n} J_{r}\cos{(rk)},
\end{align}
where $k$ is the lattice wave-vector and its range is given by $-\pi \leq k \leq \pi$. 

For studying transport we briefly introduce the notion of transfer matrices for lattice which will play an instrumental role in providing analytical insights into the behaviour of conductance \cite{subdiffusive_madhumita,hypersurface_madhumita,isolated_draft}. 
Transfer matrices are essentially a map that connect amplitudes of the single particle wavefunction $|\psi\rangle$ at different lattice sites in a recursive manner \cite{Luca_Molinari_1997,Luca_Molinari_2003,Transfer_matrix_Last,Transfer_matrix_3}. One starts with the time-independent Schr\"odinger equation $h_S |\psi \rangle = \omega |\psi \rangle$ for the single particle lattice Hamiltonian $h_S$ corresponding to Eq.~\eqref{finite-range-Ham} and then recasts it in the form of a matrix equation. More precisely, from Ref.~\onlinecite{hypersurface_madhumita}, the transfer matrix for a setup with range of hopping $n$ is given by
\begin{center}
\begin{align}
\label{general_transfer}
\mathbb{T}_{0}(\omega)=\begin{pmatrix} -\frac{J_{n-1}}{J_n} & -\frac{J_{n-2}}{J_n}& \ldots & -\frac{\omega}{J_n} & \ldots  & -\frac{J_{n-1}}{J_n} & -1 \\
1 & 0 & \ldots & \ldots & \ldots  & 0 & 0  \\
0 & 1 & \ldots & \ldots & \ldots  & 0 & 0  \\
\vdots & \vdots & \vdots & \vdots  & \vdots & \vdots & \vdots \\
0 & 0 & \ldots & \ldots & \ldots  & 0 & 0 \\
0 & 0 & \ldots & \ldots & \ldots  & 1 & 0 
\end{pmatrix}.
\end{align}
\end{center}

We next discuss the boundary reservoirs represented in Eq.~\eqref{Ham} as $H_L$ and $H_R$. Each reservoir consists of collection of infinite number of fermionic modes. The Hamiltonian corresponding to the left and the right reservoir is given as
\begin{align}
    H_{\alpha}=\sum_{k}\,\omega_{k \alpha} \,c^{\dagger}_{k\alpha} c_{k\alpha},\,\,\,\,\,\,\,\alpha \in \{L,R\},
    \label{bath-Ham}
\end{align}
where $\omega_{k \alpha}$ is the energy corresponding to the $k$-th mode for the $\alpha$-th reservoir and $c_{k \alpha} (c^{\dagger}_{k \alpha})$ is the annihilation (creation) operator for the $k$-th mode of the $\alpha$-th reservoir. 
The coupling between the system and the left reservoir is given by
\begin{align}
    H_{LS}= \sum_{k} \gamma_{kL} \,c^{\dagger}_{1}\,c_{kL}+h.c.
\end{align}
with $\gamma_{kL}$ being the coupling strength between the $k$-th mode of the left reservoir and the first site of the lattice 
and the coupling between system and the right reservoir is given by 
\begin{align}
    H_{RS}=\sum_{k} \gamma_{kR}\, c^{\dagger}_{N}\,c_{kR}+h.c.
\end{align}
with $\gamma_{kR}$ being the coupling strength between the $k$-th mode of the right reservoir and the $N$-th site of the lattice. Given this lattice setup with boundary reservoirs being at different chemical potentials, a nonequilibrium steady state (NESS) with finite current $I$ sets in. In the linear response regime ($\Delta \mu= \mu_L - \mu_R \to 0$), the steady state conductance $G = I/\Delta \mu$ shows different transport regimes depending on the value of the equilibrium chemical potential. For example, for chemical potential located within the lattice band, the conductance scales ballistically with system size i.e., $G \sim N^0$, for chemical potential located outside the system band, the conductance scales exponentially with system size i.e., $G \sim e^{-N/\xi}$ with $\xi$ being the localization length, and for the chemical potential corresponding to the band-edges of the system, the transport is sub-diffusive with $G \sim 1/N^2$ \cite{subdiffusive_madhumita}.
In the following, we investigate how these different transport regimes i.e., scaling of conductance with system size, get affected in presence of particle losses from lattice sites. In Sec.~\ref{sec-III}, we  first discuss the scenario when the lattice is subjected to single localized loss.

\section{Localized loss} \label{sec-III}
To mimic the localized particle loss from the lattice, we employ the local Lindblad Quantum Master Equation, given as
\begin{align}
    \frac{d\rho}{dt}=-i\big[H,\rho\big]+\gamma\, \Big[c_{m}\rho c_{m}^{\dagger}-\frac{1}{2}\{c_{m}^{\dagger} c_{m},\rho\}\Big],
    \label{density-loss}
\end{align}
where the first term represents the unitary evolution and the second term mimics particle loss from a site $m$ of the lattice with loss rate $\gamma$. We choose $m=\frac{N+1}{2}$ which corresponds to the middle site of the lattice. Here $H$ represents the total Hamiltonian of the setup and it is given in Eq.~\eqref{Ham}. $\rho$ represents the density operator of the full system i.e., the system and left, right reservoir. Given this Lindblad equation, one can write down the Keldysh action using Schwinger-Keldysh path integral formalism \cite{Sieberer_2016}, as obtained in Refs.~\onlinecite{Giamarchi_loss,Giamarchi_nonlinear,Giamarchi_Lindblad}. From the Keldysh action, one can identify the non-equilibrium Green's function (NEGF) of the lattice chain as
\begin{align}
    \mathcal{G}^{r}(\omega)=\Big[\omega \,\mathbb{I}_{N}-h_{S} - \Sigma^r_L(\omega)-\Sigma^{r}_R (\omega) + i\Gamma\Big]^{-1}, \label{GF}
\end{align}
where $h_{S}$ is the $N \times N$ single particle hamiltonian matrix of the lattice,  $\Sigma^r_{L}(\omega)$ and $\Sigma^r_R(\omega)$ are $N \times N$ matrix describing the self energy due to the left and the right reservoir, respectively. The self energy due to these two boundary reservoirs are exact and hence in general can be dependent on $\omega$. Here $\Sigma_L^r(\omega)|_{11} = \sum_{k} |\gamma_{kL}|^2 \, g^r_{kL}(\omega)$ and $\Sigma_R^r(\omega)|_{NN} = \sum_{k} |\gamma_{kR}|^2 \, g^r_{kR}(\omega)$ with $g_{k\alpha}^r(\omega)= \big[(\omega + i 0^{+}) - h_{\alpha}\big]^{-1}$ being the bare green function for the $\alpha$-th reservoir and $h_{\alpha}$ is the single particle hamiltonian for the reservoir $\alpha$. 
On the other hand, the self energy corresponding to the local loss is represented by the $N\times N$ matrix with element 
\begin{align}
    \Gamma_{ij}=\frac{\gamma}{2} \, \delta_{im} \, \delta_{jm},
    \label{loss-gamma}
\end{align}
where recall that $m=\frac{N+1}{2}$ corresponds to the middle site of the lattice.  Note that Eq.~\eqref{loss-gamma} can be systematically obtained starting from the Lindbladian form given in Eq.~\eqref{density-loss} without any further approximation \cite{Giamarchi_Lindblad}. 

We next discuss the approximation of wide-band limit for the boundary reservoirs. This limit involves assuming bath spectral function to be independent of $\omega$ which subsequently yields, 
\begin{equation}
    \Sigma_L^r(\omega)|_{11} = -i \gamma_L/2 \,\,\,\,\, {\rm {and}} \, \,\, \,\, \Sigma_R^r(\omega)|_{NN} = -i \gamma_R/2,
    \label{wide-band}
\end{equation}
and rest of the matrix elements are zero. Thus the self-energies that enter Eq.~\eqref{GF}, are Eq.~\eqref{loss-gamma} and Eq.~\eqref{wide-band} for the loss channel and the reservoirs, respectively.

We now discuss our primary quantities of interest, namely, the current and the conductance. Having obtained the Green's function [Eq.~\eqref{GF}], the net current drawn from the left reservoir is given by
\begin{align}
    I_{L}&=I_{L}^{0}+I_{L}^{1}\nonumber\\
    &=\int_{-\infty}^{\infty} \frac{d\omega}{2\pi} \, T(\omega)(f_{L}(\omega)-f_R(\omega))+\mathcal{L}_{L}(\omega)f_{L}(\omega),
    \label{left_current}
\end{align}
where $I_{L}^{0}$ is the current coming out from the left reservoir and entering to the right reservoir and $I_{L}^{1}$ is the current coming out from the left reservoir but lost through the dissipative channel located at site $m$ (see Fig.~\ref{schematic-local-loss}). Here 
$T(\omega)=\gamma_{L}\gamma_{R}|\mathcal{G}^{r}_{1N}(\omega)|^{2}$ is the transmission function from the left reservoir to the right reservoir and $\mathcal{L}_{L}(\omega)=\gamma\gamma_{L}|\mathcal{G}^{r}_{1m}(\omega)|^{2}$ is the transmission function characterizing the loss from the left reservoir to the dissipative site $m$. In Eq.~\eqref{left_current},  $f_{\alpha}(\omega)=\big[1+e^{\beta (\omega - \mu_{\alpha})}\big]^{-1}$ is the Fermi function for the $\alpha$-th boundary reservoir with chemical potential $\mu_{\alpha}$ and inverse temperature $\beta$. Similarly, one can write down the net current entering into the right reservoir as (see Fig.~\ref{schematic-local-loss}),
\begin{align}
    I_{R}&=I_{R}^{0}-I_{R}^{1}\nonumber\\
    &=\int_{-\infty}^{\infty} \frac{d\omega}{2\pi} T(\omega)(f_{L}(\omega)-f_R(\omega))-\mathcal{L}_{R}(\omega)f_{R}(\omega),
    \label{right_current}
\end{align}
where $I_R^{0}$ is the current coming into the right reservoir from the left reservoir and is equal to $I_L^{0}$. $I_R^{1}$ is the current that is drawn from the right reservoir and that flows through the dissipated site $m$. Therefore, the net current entering into the right reservoir is the difference between $I_R^{0}$ and $I_R^{1}$.  In Eq.~\eqref{right_current}, $\mathcal{L}_{R}(\omega)=\gamma\gamma_R|\mathcal{G}^{r}_{Nm}(\omega)|^{2}$ is the transmission function characterizing the loss from the right reservoir to the dissipated site $m$. For the case of $\gamma_L=\gamma_R=\gamma_0$ and the lossy channel assumed to be in the middle, we have  $\mathcal{L}_{L}(\omega)=\mathcal{L}_{R}(\omega)= \mathcal{L}(\omega)$. Therefore from Fig.~\ref{schematic-local-loss}, and Eqs.~\eqref{left_current} and \eqref{right_current}, one can find that the net current flowing through the system from left to right [i.e., in the direction of bias ($\mu_L>\mu_R$)] is given by \cite{Uchino_loss,uchino2023particlecurrentnoisecounting},
\begin{align}
    I&=\frac{I_L + I_R}{2} = \frac{I_{L}^{0}+I_{L}^{1}+I_{R}^{0}-I_{R}^{1}}{2}\nonumber\\
    &=\int_{-\infty}^{+\infty}\frac{d\omega}{2\pi}\Big[T(\omega)+\frac{\mathcal{L}(\omega)}{2}\Big]\big(f_{L}(\omega)-f_{R}(\omega)\big). 
    \label{left-right-current}
\end{align}
Next we consider the linear response and zero temperature limit in Eq.~\eqref{left-right-current} for the left and right reservoirs and obtain the exact conductance $G=I/\Delta\mu$ as,
\begin{align}
G(\mu)=\Big[T(\mu)+\frac{\mathcal{L}(\mu)}{2}\Big].\label{conductance}
\end{align}
Here $\mu=\frac{\mu_L+\mu_R}{2}$ and $\Delta\mu=(\mu_L-\mu_R)\rightarrow0
 $ in the linear response limit and we recall that 
 \begin{eqnarray}
 \label{T-L}
     T(\mu) &=&\gamma_0^{2}|\mathcal{G}^{r}_{1N}(\mu)|^{2}, \\
     \mathcal{L}(\mu)&=&\gamma\gamma_0|\mathcal{G}^{r}_{1m}(\mu)|^{2}.
     \label{L-mu}
 \end{eqnarray} 
 In this work, we are interested in understanding the behavior of the conductance $G(\mu)$, primarily its scaling with system size $N$, in the presence of loss. We consider the case when the chemical potential $\mu$ is located within, outside, or at the band edge of the lattice. In particular, we show cases when the transport displays  non-trivial behavior when $\mu$ is located at the band-edges, the cause of which is rooted in the interplay between loss channel and the extrema nature of the band edges. In what follows, we first discuss the situation when the lattice has only the nearest neighbor hopping ($n=1$) term with hopping amplitude $J$. 

\vspace{1cm}
\subsection{Conductance scaling in presence of loss for $n=1$}
\label{sub_1}
We now discuss the case of $n=1$ which is the standard tight-binding model with nearest neighbor hopping. It turns out even when this system is subjected to boundary reservoirs and lossy channel, it is still amenable to analytical calculations. In order to calculate the conductance following Eq.~\eqref{conductance}, the matrix elements of the NEGF in Eq.~\eqref{GF} needs to be computed. For $n=1$, given the tridiagonal structure of the single particle hamiltonian $h_S$, calculating the inverse of ${\cal G}^r(\mu)$ involves the inversion of a tridiagonal matrix which yields \cite{Abhishek_heat}
 \begin{align}
     &\mathcal{G}^{r}(\mu)=\!-\frac{1}{J}\big[M(\mu)\big]^{-1},\label{GF_element} \\
     & M(\mu)\!=\!-\frac{1}{J}\big[\mu \mathbb{I}_N\!-h_{S}\! + i \Gamma_0 + i \Gamma\big], \label{M-inverse}\\
     &\big[M(\mu)\big]_{ij}^{-1}=(-1)^{i+j}\frac{\Delta_{1,i-1}(\mu)\Delta_{j+1,N}(\mu)}{\Delta_{1,N}(\mu)}. \label{M_element}
 \end{align}
Here $\Delta_{l,k}$ is the determinant of the sub-matrix starting with $l$-th row and $l$-th column and ending with $k$-th row and $k$-th column of the matrix $M(\mu)$.
Here $\Gamma_0|_{11} = \Gamma_0|_{NN} = \gamma_0/2$ and zero otherwise. Interestingly, the determinants that appear in Eq.~\eqref{M_element} can be expressed in terms of the transfer matrix of the lattice [see Eq.~\eqref{general_transfer}],
\begin{widetext}
\begin{align}
    &\begin{pmatrix}
            \Delta_{1,N}(\mu) \\
            \Delta_{2,N}(\mu) \\
        \end{pmatrix}=\begin{pmatrix}
            1 & \,\,\,\,-\frac{i\gamma_0}{2J} \\
            0 & 1 \\ 
        \end{pmatrix}\Big[\mathbb{T}_{0}(\mu)\Big]^{\frac{N-1}{2}}\begin{pmatrix}
            -\frac{\mu}{J}-\frac{i\gamma}{2J} &\,\,\,\, -1 \\
            1  & 0 \\
        \end{pmatrix}\Big[\mathbb{T}_{0}(\mu)\Big]^{\frac{N-1}{2}}\begin{pmatrix}
            1 \\
            \frac{i\gamma_0}{2J}
        \end{pmatrix}, \label{T1}
\end{align}
\begin{align}
        &\begin{pmatrix}
            \Delta_{1,l-1}(\mu) \\
            \Delta_{2,l-1}(\mu) \\
        \end{pmatrix}=\begin{pmatrix}
            1 & \,\,\,\,-\frac{i\gamma_0}{2J} \\
            0 & 1 \\ 
        \end{pmatrix}\Big[\mathbb{T}_{0}(\mu)\Big]^{\frac{N-1}{2}}\begin{pmatrix}
           -\frac{\mu}{J}-\frac{i\gamma}{2J} & \,\,\,\,-1 \\
            1  & 0 \\
        \end{pmatrix}\Big[\mathbb{T}_{0}(\mu)\Big]^{l-\frac{N+3}{2}}\begin{pmatrix}
            1 \\
           0 \\
        \end{pmatrix}\,\,\,(\mathrm{for}\,\,l-1\ge m),\label{T2}
\end{align}
\begin{align}
        &\begin{pmatrix}
            \Delta_{1,l-1}(\mu) \\
            \Delta_{2,l-1}(\mu) \\
        \end{pmatrix}=\begin{pmatrix}
            1 & -\frac{i\gamma_0}{2J} \\
            0 & 1 \\ 
        \end{pmatrix}\Big[\mathbb{T}_{0}(\mu)\Big]^{l-1}\begin{pmatrix}
            1 \\
           0 \\
        \end{pmatrix}\,\,\,\,\,\,(\mathrm{for}\,\,l-1<m)\label{T3}
\end{align}
\begin{align}
        &\begin{pmatrix}
            \Delta_{l,N}(\mu) \\
            \Delta_{l+1,N}(\mu) \\
        \end{pmatrix}=\Big[\mathbb{T}_{0}(\mu)\Big]^{N-l+1}\begin{pmatrix}
            1 \\
           \frac{i\gamma_0}{2J} \\
        \end{pmatrix}\,\,\,\,\,\,(\mathrm{for}\,\,l>m)\label{T4}
\end{align}
\begin{align}
        &\begin{pmatrix}
            \Delta_{l,N}(\mu) \\
            \Delta_{l+1,N}(\mu) \\
        \end{pmatrix}=\Big[\mathbb{T}_0(\mu)\Big]^{\frac{N+1}{2}-l}\begin{pmatrix}
            -\frac{\mu}{J}-\frac{i\gamma}{2J} & \,\,\,\,-1 \\
            1  & 0 \\
        \end{pmatrix}\Big[\mathbb{T}_{0}(\mu)\Big]^{\frac{N-1}{2}}\begin{pmatrix}
            1 \\
           \frac{i\gamma_0}{2J} \\
        \end{pmatrix}\,\,\,\,\,\,(\mathrm{for}\,\,l\le m).\label{T5}
    \end{align}
\end{widetext}
Here in Eqs.~\eqref{T1}-\eqref{T5}, $\mathbb{T}_{0}(\mu)$ is the $2\times 2$ transfer matrix (since $n=1$), given by \cite{subdiffusive_madhumita}
\begin{equation}
\label{T0}
\mathbb{T}_{0}(\mu)=\begin{pmatrix} -\frac{\mu}{J} & -1 \\ 1 & 0\\
\end{pmatrix}, \quad \quad {\rm for} \,\, \, n=1. 
\end{equation}
Given the prescription to calculate ${\cal G}^r_{ij}(\mu)$ in Eq.~\eqref{GF_element}, we can now investigate the consequence of loss on transport when $\mu$ is located within, outside or at the band-edge of the lattice. In particular, we obtain the required Green's function elements $\mathcal{G}^{r}_{1N}(\mu)$ and $\mathcal{G}^{r}_{1m}(\mu)$ to compute the conductance in presence of localized loss.  For $n=1$, the dispersion relation corresponding to the single-particle hamiltonian is,
\begin{align}
    \omega(k)=-2 J \cos(k), \quad -\pi \leq k \leq \pi, \quad {\rm for} \, \, n=1
    \label{dispersion-nn}
\end{align}
which implies that $|\mu|< 2J$ ($|\mu|>2J$) corresponds to chemical potential lying within (outside) the band, and $|\mu|=2J$ corresponds to the band-edges. 

{\it Within the band --} When $\mu$ is fixed anywhere inside the band i.e., $|\mu|<2J$, the transfer matrix $\mathbb{T}_{0}(\mu)$ for $n=1$ in Eq.~\eqref{T0} is diagonalizable and the eigenvalues are complex conjugate pairs with absolute value 1 i.e., $\lambda_{\pm}=e^{\pm i \kappa}$ where $\kappa$ is the real solution for $\mu = -2 J \cos(\kappa)$. Hence, from Eq.~\eqref{T1} and Eq.~\eqref{T5}, we obtain,
\begin{align}
    &\Delta_{1,N}(\mu)\propto e^{-i \kappa N}, \,\,\,\,\,\,
    \Delta_{m+1,N}\propto e^{-i \kappa m},\label{Delta-within-band}
\end{align}
where recall that $m=\frac{N+1}{2}$ is the central site of the lattice at which the loss is present.
Using Eq.~\eqref{Delta-within-band}, we obtain the Green's function components as,
\begin{align}
    &|\mathcal{G}^{r}_{1N}(\mu)|^{2}=\frac{1}{|\Delta_{1,N}(\mu)|^{2}}\propto N^0, \label{GF-element-1N-within-band} \\
    &|\mathcal{G}^{r}_{1m}(\mu)|^{2}=\frac{|\Delta_{m+1,N}(\mu)|^{2}}{|\Delta_{1,N}(\mu)|^{2}}\propto N^0. \label{GF-element-1m-within-band}
\end{align}
Hence the conductance $G(\mu)$ in Eq.~\eqref{conductance} for any $\mu$ located within the band scales as,
\begin{align}
    G(\mu)=\Big[\gamma_{0}^{2} \,|\mathcal{G}^{r}_{1N}(\mu)|^{2}+\frac{\gamma\gamma_0}{2}\,|\mathcal{G}^{r}_{1m}(\mu)|^{2}\Big] \propto N^{0}. \label{conductance-within-band}
\end{align}
This implies that the conductance is independent of $N$, i.e., its scaling with system size is $N^{0}$. 

{\it Outside the band --} On the other hand, when $\mu$ is located somewhere outside the band ($|\mu|>2J$), the transfer matrix 
$\mathbb{T}_{0}(\mu)$ in Eq.~\eqref{T0} is once again diagonalizable but the eigenvalues of the transfer matrix are now real i.e., $\lambda_{\pm}=e^{\pm i \kappa }$ where $\kappa$ is the purely imaginary solution for $\mu=-2 J  \cos(\kappa)$. Thus we find,
\begin{align}
    \Delta_{1,N}(\mu)\propto e^{|\kappa| N},\,\,\,\,
    \Delta_{m+1,N}\propto e^{|\kappa| m}.\label{Delta-outside-band}
\end{align}
Using Eq.~\eqref{Delta-outside-band}, we obtain the Green's function elements as,
\begin{align}
    &|\mathcal{G}^{r}_{1N}(\mu)|^{2}=\frac{1}{|\Delta_{1,N}(\mu)|^{2}}\propto e^{-|\kappa| N}\label{GF-element-1N-outside-band}\\
    &|\mathcal{G}^{r}_{1m}(\mu)|^{2}=\frac{|\Delta_{m+1,N}(\mu)|^{2}}{|\Delta_{1,N}(\mu)|^{2}}\propto e^{-\frac{|\kappa| N}{2} } \label{GF-element-1m-outside-band}.
\end{align}
Thus, in presence of local loss, outside the band the conductance scales exponentially with the system size i.e., $G(\mu)\sim e^{-|\kappa| N}$ with localization length $\propto 1/|\kappa|$.  Therefore, we find that the conductance scaling remains unaltered even in the presence of local loss when $\mu$ is located either within the band or outside the band. However, this is not the case when $\mu$ is located at the band edge,  in which case the presence of local loss yields markedly different transport behavior. 
 
{\it At the band-edge --} We now investigate the impact of local loss at the band-edge i.e., at $|\mu|= 2J$. Interestingly, the transfer matrix $\mathbb{T}_{0}(\mu)$ in Eq.~\eqref{T0} at $\mu=\pm 2J$ is non-diagonalizable as both the two eigenvalues and the eigenvectors coalesce and thus it is a second order exceptional point (EP) \cite{subdiffusive_madhumita,EP_chen,EP_wiersig,EP_Zhong,EP_Hashemi,EP_Mandal}. The transfer matrix at this EP can be transformed to a Jordan normal form. Let us consider $\mu=2J $ case i.e., the upper band edge. In this case, the Jordan normal form $R$ is given as,
\begin{align}
    R=S\mathbb{T}_{0}(\mu)S^{-1}=\begin{pmatrix}
    -1 & \,1\\
    0 & -1\\
\end{pmatrix},\,\,\,\mathrm{where}\,\,
S=\begin{pmatrix}
    0 & 1 \\
    1 & 1\\
\end{pmatrix}. \label{Jordan-normal}
\end{align}
Here the coalesced eigenvalues of the transfer matrix are $\lambda= -1$. Using this Jordan normal form in Eq.~\eqref{Jordan-normal} and following Eqs.~\eqref{T1}-\eqref{T5}, we calculate the Green's function elements at the upper band-edge. The expressions for $\Delta_{1,N}(\mu)$ and $\Delta_{m+1,N}(\mu)$ for $\mu=2J$ are given as (henceforth setting $\gamma_0=J=1$ which sets the energy scale),
\begin{align}
    \Delta_{m+1,N}(\mu) &= \frac{1}{4}(-1)^{\frac{N-1}{2}}\Big[2(N+1)+i(N-1)\Big],
\end{align}
\begin{align}
    \Delta_{1,N}(\mu)\! = \frac{1}{32}&\Big[2i(N+1)-(N-1)\Big]\times \nonumber\\& \Big[2i\gamma(N+1)-\gamma (N-1)+8i+16\Big]. \label{delta-exact-1}
\end{align}
We now discuss the emergence of different transport regimes at the band-edge by analysing the system size scaling of conductance and also provide an estimate of the corresponding transport windows. 
Using Eq.~\eqref{delta-exact-1} and assuming $\gamma \ll 1$, we obtain
\begin{equation}
|\Delta_{1,N}(\mu)|^{2} \sim
    \begin{cases} 
\frac{25}{16}N^{2} \quad \quad \quad{\mathrm {for}} \, \, N< B_2 \, \mathcal{O}(\frac{1}{\gamma})\\ \\
\frac{25}{1024}\gamma^{2} N^{4}   \quad \quad {\mathrm {for}} \,\, N>  B_2 \, \mathcal{O}(\frac{1}{\gamma})
\end{cases} 
\label{Delta_1_N_n1}
\end{equation}
where the pre-factor can be estimated to be $B_2 \approx  8$.

We would like to remark that in equations such as Eq.~\eqref{Delta_1_N_n1} one typically provides just an order estimate such as $\mathcal{O}(1/\gamma)$. However, we were able to provide more accurate estimates reflected by the pre-factors such as $B_2$ in Eq.~\eqref{Delta_1_N_n1}. In other cases as well, as will be seen later, we were able to estimate the pre-factors.

Thus for small values of $\gamma$, there is a range of $N$, $N< B_2 \, \mathcal{O}(\frac{1}{\gamma})$, for which the Green's function elements in Eq.~\eqref{GF_element} scales as,
\begin{align}
    &|\mathcal{G}^{r}_{1N}(\mu)|^{2}=\frac{1}{|\Delta_{1,N}(\mu)|^{2}}\sim \frac{16}{25} \frac{1}{N^{2}},\label{GF-element-1N-band-edge}\\
    &|\mathcal{G}^{r}_{1m}(\mu)|^{2}=\frac{|\Delta_{m+1,N}(\mu)|^{2}}{|\Delta_{1,N}(\mu)|^{2}}\sim \frac{1}{5}N^{0}. \label{GF-element-1m-band-edge}
\end{align}
Therefore, the conductance $G(\mu)$ defined in Eq.~\eqref{conductance} is,
\begin{align}
    G(\mu) \sim \Big[\frac{16}{25}\frac{1}{N^{2}}+\frac{\gamma}{10}N^0\Big] \label{cond-approx-n1}
\end{align}
As a result, for $N< A_2 \, \mathcal{O}(1/\gamma^{1/2})$ where $A_2 \approx 3$, the conductance $G(\mu)\propto 1/N^{2}$ (subdiffusive) and for the range $ A_2 \, \mathcal{O}(1/\gamma^{1/2})<N<B_2 \, \mathcal{O}(1/\gamma)$, the conductance scales as, $G(\mu)\propto N^{0}$ (ballistic plateau).

Finally for $N>B_2\,\mathcal{O}(1/\gamma)$, 
the Green's function elements become,
\begin{align}
    &|\mathcal{G}^{r}_{1N}(\mu)|^{2}\sim \frac{1024}{25}\frac{1}{\gamma^{2}N^{4}},\label{GF-1N-large-N}\\
    &|\mathcal{G}^{r}_{1m}(\mu)|^{2}\sim  \frac{128}{5}\frac{1}{\gamma^2 N^{2}}.\label{GF-1m-large-N}
\end{align}
Therefore for $N > B_2\,\mathcal{O}(1/\gamma)$ the conductance scales as $ G(\mu) \propto  1/N^{2}$.

\begin{figure}[h!]
     \centering
     \includegraphics[width=8.5cm]{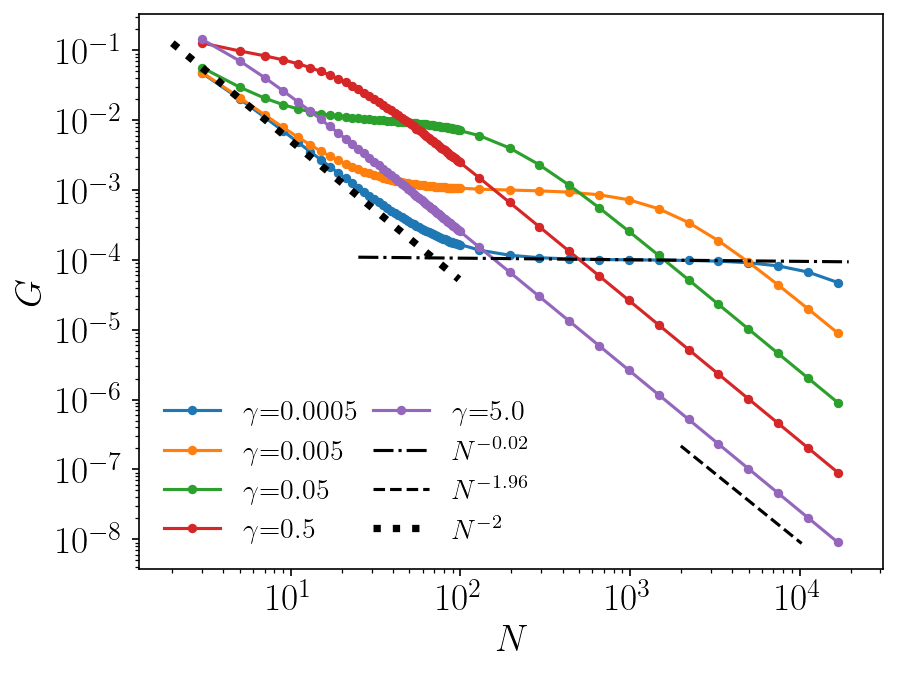}
     \caption{Scaling of exact conductance $G(\mu)$ given in Eq.~\eqref{conductance} with  system size $N$ at the upper band-edge ($\mu=2J)$ for the nearest neighbor hopping model ($n=1$) for different values of the loss strength $\gamma$. Here $J=\gamma_0=1$. The different transport regimes namely first subdiffusive regime, followed by a ballistic plateau and finally the second subdiffusive regime are in agreement with those predicted in Eq.~\eqref{G-n1}.}
     \label{2nd-order-local}
 \end{figure}

To summarize, for $n=1$, the conductance at the band-edges has interesting subdiffusive-ballistic-subdiffusive characteristics as we increase the system size $N$. In other words the dominant system size contribution to conductance can be summarized as,
\begin{equation}
\label{G-n1}
   G(\mu)  \propto 
\begin{cases} 
\frac{1}{N^2} \quad \quad {\mathrm {for}} \, \, \,\, N< A_2 \, \mathcal{O}\Big(\frac{1}{\gamma^{1/2}}\Big)  \\ \\
 N^{0}    \quad \quad {\mathrm {for}} \, \, \, \,  A_2 \, \mathcal{O}\Big(\frac{1}{\gamma^{1/2}}\Big)< N< B_2 \, \mathcal{O}\Big(\frac{1}{\gamma}\Big)  \\ \\
\frac{1}{N^2}   \quad \quad {\mathrm {for}} \, \,\, \,  N> B_2 \, \mathcal{O}\Big(\frac{1}{\gamma}\Big)
\end{cases} 
\end{equation}
where the constants $A_2$ and $B_2$ can be estimated and given as $A_2 \approx 3 $, $B_2 \approx 8$. These anomalous transport regimes are clearly demonstrated in Fig.~\ref{2nd-order-local} where we show the conductance scaling as a function of $N$ for different values of loss strength $\gamma$. We observe a clear $1/N^2$ regime for $N < A_{2}\, \mathcal{O}(1/\gamma^{1/2})$. After this, the ballistic plateau emerges and this trend continues upto $N < B_2\,\mathcal{O}(1/\gamma)$. For $N>B_2\,\mathcal{O}(1/\gamma)$, we once again observe a clear subdiffusive regime with $1/N^2$ scaling. The location where the trend changes precisely matches with our analytical predictions in Eq.~\eqref{G-n1}. Our analysis shows that one can appropriately control these various regimes of transport by appropriately tuning $\gamma$. It is important to note that, the first subdiffusive regime with scaling $1/N^2$ emerges due to the term $|{\mathcal G}_{1N}^r(\mu)|^2$, whereas the second subdiffusive regime with the same scaling  emerges due to the local loss term $|\mathcal{G}^{r}_{1m}(\mu)|^{2}$. 
 \begin{figure}[h!]
    \centering
    \includegraphics[width=8.3cm]{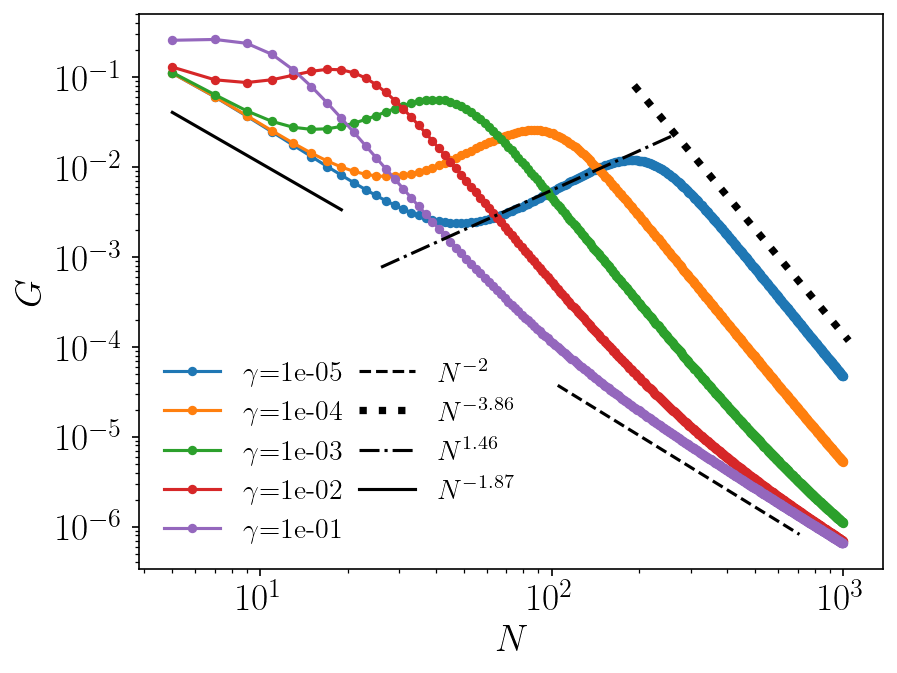}
    \caption{Scaling of exact conductance $G(\mu)$ given in Eq.~\eqref{conductance} with system size $N$ at the upper band-edge corresponding to $\mu = 6J_2$ for range of hopping $n=2$ for different values of loss strength $\gamma$. The upper band-edge in this case hosts a fourth order exceptional point. Here $\gamma_0=1$ and $J_1=1$ and $J_2=1/4$. There is a crossover in transport regime from first subdiffusive to superballistic to second subdiffusive to a third subdiffusive regime. The analytical arguments for the observed behaviour is presented in detail in Appendix \ref{appendix-A}.}
    \label{4th_order_local}
\end{figure}
 Note that for such nearest neighbour hopping ($n=1$) lattice setup, the transport at the band-edge ($\mu = \pm 2J$) can utmost be enhanced to the ballistic regime by subjecting the particle loss. The emergence of this ballistic regime is due to the existence of a second  order exceptional point which is the highest order EP for $n=1$. Since the transport properties can be sensitive to the order of the EP \cite{saha2024effectordertransfermatrix}, in what follows we focus on lattices with range of hopping $n>1$. For such cases, the transfer matrix hosts higher order (beyond second order) EP and we show that this feature subsequently leads to rich transport behaviour at the band edges in presence of loss. However, it can be shown that the conductance scaling $G(\mu)$ within and outside the band always remains unaltered in presence of local loss and therefore insensitive to the order of the EP. 
 
 \subsection{Conductance scaling in presence of loss for $n > 1$}
In this subsection, we discuss the conductance scaling at the band-edges in presence of local loss when one has order of EP greater than two. This can be realized by introducing hopping processes beyond nearest neighbour [see Eq.~\eqref{finite-range-Ham}].  

For a finite range hopping lattice with range of hopping $n$, the transfer matrix $\mathbb{T}_0(\mu)$ in Eq.~\eqref{general_transfer} is a $2n \times 2n$ matrix and as a result can at most host $2n$-th order exceptional point. Such highest order ($2n$-th) EPs can be systematically generated at the lower and the upper band edges corresponding to $k=0$ and $k=\pm \pi$, respectively. This can be achieved by demanding  \cite{saha2024effectordertransfermatrix, isolated_draft} 
\begin{align}
    \frac{d^{s}\omega(k)}{dk^{s}}\Bigg|_{k_{b}}=0,\,\,\,\, s=1,2,\dots,2n-1, \label{condition-EP}
\end{align}
where recall that $\omega(k)$ is the dispersion relation of the finite range lattice, as given in Eq.~\eqref{dispersion} and $k_{b}$ corresponds to band-edge. Recently, the impact of such higher order EPs that occur at the band edges, on conductance scaling was studied in absence of loss \cite{hypersurface_madhumita, saha2024effectordertransfermatrix}. It was shown that the transport at the band edges is always subdiffusive with universal $1/N^2$ scaling and independent of the order of the EP. Here we analyse how the transport scaling changes for such higher order EPs when the lattice is subjected to local loss. 

\begin{center}
\begin{table*}
    \begin{tabularx}{1.0\textwidth} { 
  | >{\centering\arraybackslash}X 
  | >{\centering\arraybackslash}X 
  | >{\centering\arraybackslash}X
  | >{\centering\arraybackslash}X 
  | >{\centering\arraybackslash}X 
  | >{\centering\arraybackslash}X | }
 \hline
 Order of the EP $(p)$ & Early $N$ & 1st intermediate $N$ & 2nd intermediate $N$ & Late $N$ & Figures\\
 \hline
 2   & $1/N^{2}$ & $N^0$ & $1/N^{2}$ & $1/N^{2}$ & Fig.~\ref{2nd-order-local}\\
 \hline
 4  & $1/N^{2}$  & $N^{2}$& $1/N^{4}$ & $1/N^{2}$ & Fig.~\ref{4th_order_local}\\
\hline
 6  & $1/N^{3}$ & $N^{4}$ & $1/N^{6}$& $1/N^{2}$ & Fig.~\ref{high-order-EP-local}(a)\\
 \hline
 8  & $1/N^{4}$ & $N^{6}$& $1/N^{8}$ & $1/N^{2}$ & Fig.~\ref{high-order-EP-local}(b)\\
 \hline
 10 & $1/N^{5}$ & $N^{8}$ & $1/N^{10}$ & $1/N^{2}$ & Fig.~\ref{high-order-EP-local}(c) \\
 \hline
\end{tabularx}
\caption{Different transport regimes at the band-edge based on conductance [Eq.~\eqref{conductance}] scaling with the system size $N$ for different order of exceptional point in presence of local loss at the middle site of the lattice (Sec.~\ref{sec-III}). The different orders of exceptional point at the band-edges can be engineered by tuning the hopping parameters of the finite range hopping model given in Eq.~\eqref{finite-range-Ham} where $n$ denotes the range of hopping.}
\label{EP-table}
\end{table*}
\end{center}

\subsubsection{n=2 case}
We first discuss the $n=2$ case.  Following the condition in Eq.~\eqref{condition-EP}, a fourth order EP can be generated at the upper band edge $k_{b}=\pm \pi$ for $J_1=4J_2$ \cite{hypersurface_madhumita}. The corresponding chemical potential value is given by $\mu = 3 J_1/2$. For this value of $\mu$, in Fig.~\ref{4th_order_local} we present the conductance scaling as a function of system size for different loss strength $\gamma$. After the initial subdiffusive $1/N^2$ scaling,  we observe an interesting superballistic transport regime. This regime is markedly distinct from the corresponding behaviour for $n=1$ where a ballistic plateau was reported [recall Fig.~\ref{2nd-order-local}]. 
Beyond this superballistic scaling regime, we observe re-emergence of rich subdiffusive behavior. We find that the exponent of the subdiffusive scaling is $1/N^4$ for a significant window of system size $N$ after which we get $1/N^2$ subdiffusive scaling for large $N$. The appearance of different transport regimes and the corresponding system size scaling in Fig.~\ref{4th_order_local}
can be analytically explained by obtaining the NEGF $\mathcal{G}^{r}(\mu)$ in Eq.~\eqref{GF_element} only in the presence of loss channel
and ignoring the self energy due to the boundary reservoirs i.e., setting $\Gamma_0=0$. We denote this Green's function as $g^r(\mu)$.  The conductance expression in Eq.~\eqref{conductance} then reduces to 
\begin{align}
    G(\mu)=\Big[\gamma_{0}^{2} \,|{g}^{r}_{1N}(\mu)|^{2}+\frac{\gamma\gamma_0}{2}\,|{g}^{r}_{1m}(\mu)|^{2}\Big] 
    \label{conductance-within-band-app-gr-1}
\end{align}
and is correct upto $\mathcal{O}(\gamma_0)$. We write $g^{r}(\mu)$ as 
\begin{align}
    g^{r}(\mu)=\big[\mu\mathbb{I}_{N}-h_{S}+i\Gamma\big]^{-1} \label{Mmu}.
\end{align}
The required elements of $g^{r}(\mu)$ can be obtained in  following the transfer matrix technique and are given by [see Appendix \ref{appendix-A}], 
\begin{align}
\label{G-1N-n-2-1}
  |g_{1N}^{r}(\mu)|^{2} \sim
\begin{cases}
\frac{4}{N^2}\,\,\,\mathrm{for}\,\,N<B_4\,\mathcal{O}\Big(\frac{1}{\gamma^{1/3}}\Big) \\
    \frac{1}{N^2}\,\,\,\mathrm{for}\,\,N>B_4\,\mathcal{O}\Big(\frac{1}{\gamma^{1/3} }\Big)
\end{cases}
\end{align}
where $B_4 \approx 6$. For the other Green's function element we obtain 
\begin{align}
\label{G-1m-n-2-2}
    |g_{1m}^{r}(\mu)|^{2} \sim
\begin{cases}
    \frac{N^2}{64}\,\,\,\mathrm{for}\,\,N<B_4\,\mathcal{O}\Big(\frac{1}{\gamma^{1/3}}\Big)\\
    \frac{144}{\gamma^2 N^{4}}\,\,\,\mathrm{for}\,\,N>B_4\,\mathcal{O}\Big(\frac{1}{\gamma^{1/3}}\Big) 
\end{cases}
\end{align}
Remarkably, due to the presence of fourth order EP at the band-edge, $|g^{r}_{1m}|^2$ in Eq.~\eqref{G-1m-n-2-2} scales as $N^2$ for $N<\mathcal{O}(1/\gamma^{1/3})$. Recall that this is in stark contrast to the case of a the second order EP $(n=1)$ where  $|g^{r}_{1m}|^2 \propto N^{0}$. Eqs.~\eqref{G-1N-n-2-1} and \eqref{G-1m-n-2-2} can be used to obtain all the different transport regime for conductance by following Eq.~\eqref{conductance-within-band-app-gr}.

Here we summarize the dominant contribution to the scaling of conductance and the corresponding transport window for $n=2$, 
\begin{align}
\label{G-mu-n-2}
G(\mu) \propto
    \begin{cases}
        \frac{1}{N^{2}}\quad \mathrm{for}\,\, N<A_4\,\mathcal{O}\Big(\frac{1}{\gamma^{1/4}}\Big) \\ \\
        N^{2} \quad \mathrm{for}\,\, A_4\,\mathcal{O}\Big(\frac{1}{\gamma^{1/4}}\Big)<N<B_4\,\mathcal{O}\Big(\frac{1}{\gamma^{1/3}}\Big) \\ \\
        \frac{1}{N^{4}} \quad \mathrm{for}\,\, B_4\,\mathcal{O}\Big(\frac{1}{\gamma^{1/3}}\Big)<N<C_4\,\mathcal{O}\Big(\frac{1}{\gamma^{1/2} }\Big)\\ \\
        \frac{1}{N^{2}}\quad \mathrm{for}\,\, N>C_4\,\mathcal{O}\Big(\frac{1}{\gamma^{1/2}}\Big)
    \end{cases}
\end{align}
where one can estimate the constants $A_4 \approx 5 $, $B_4 \approx 6$, and $C_4 \approx 9$. It is clear from Eq.~\eqref{G-mu-n-2} that the window of different transport regimes can be carefully adjusted by tuning the loss strength $\gamma$. Recall that in Eq.~\eqref{G-mu-n-2} we have set $\gamma_0=1$. The reason for providing the pre-factors can be recalled in the discussion below Eq.~\eqref{Delta_1_N_n1}.
\begin{figure*}
    \centering
    \includegraphics[width=6.0cm]{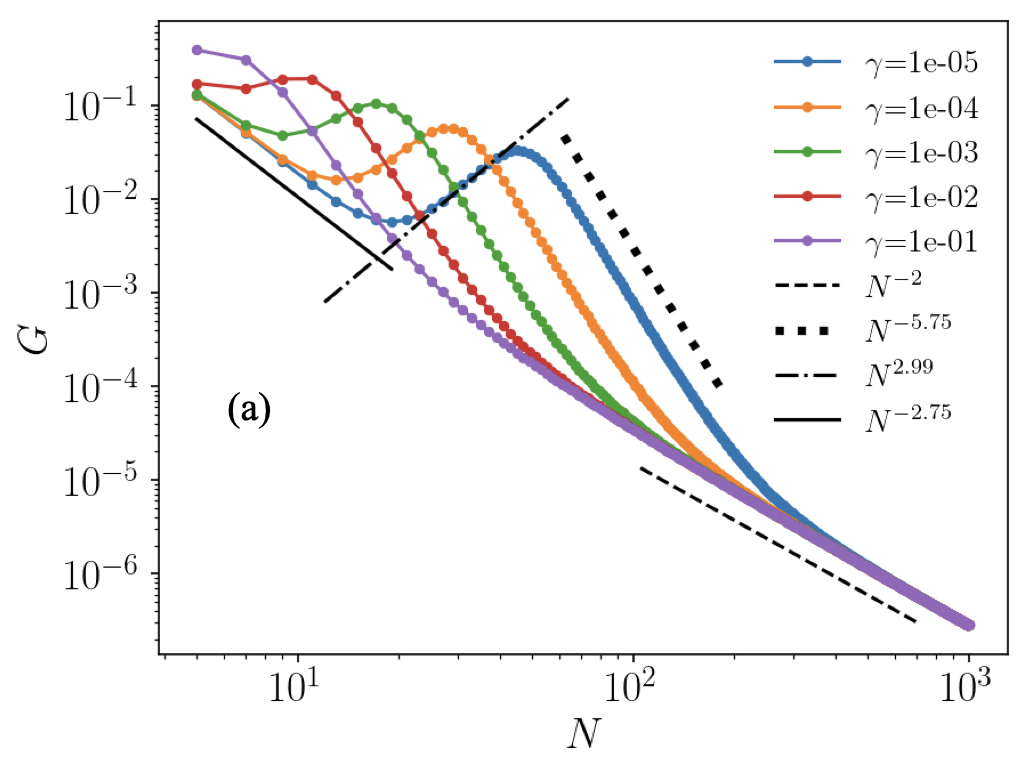}%
    \includegraphics[width=6.0cm]{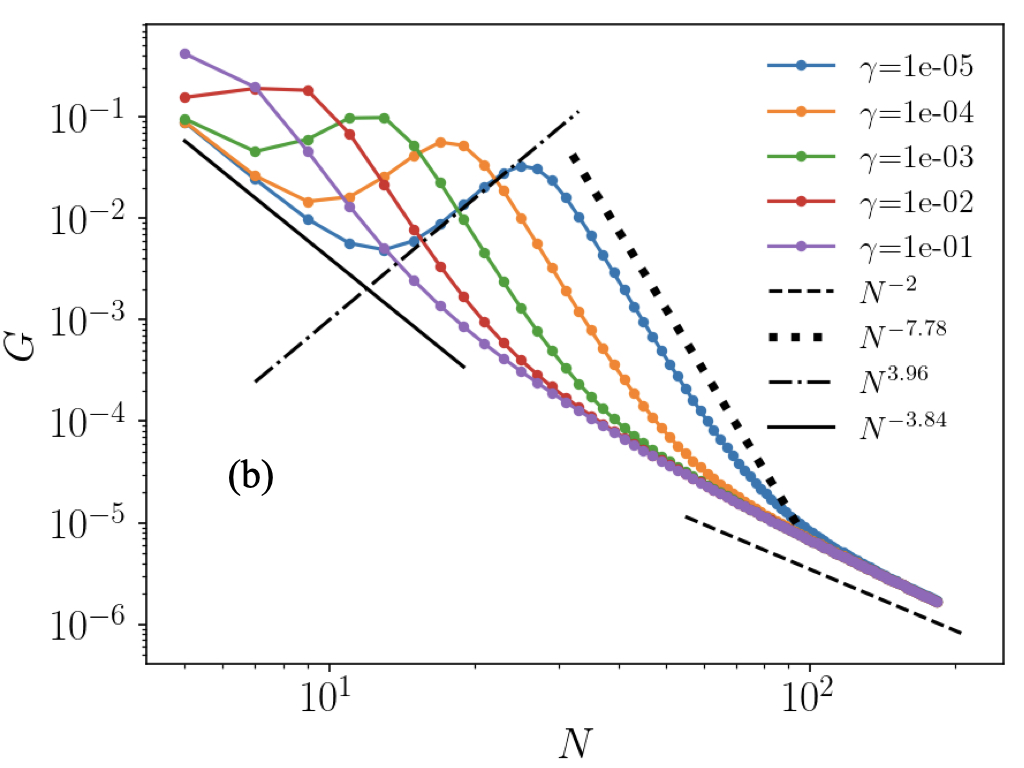}%
    \includegraphics[width=6.0cm]{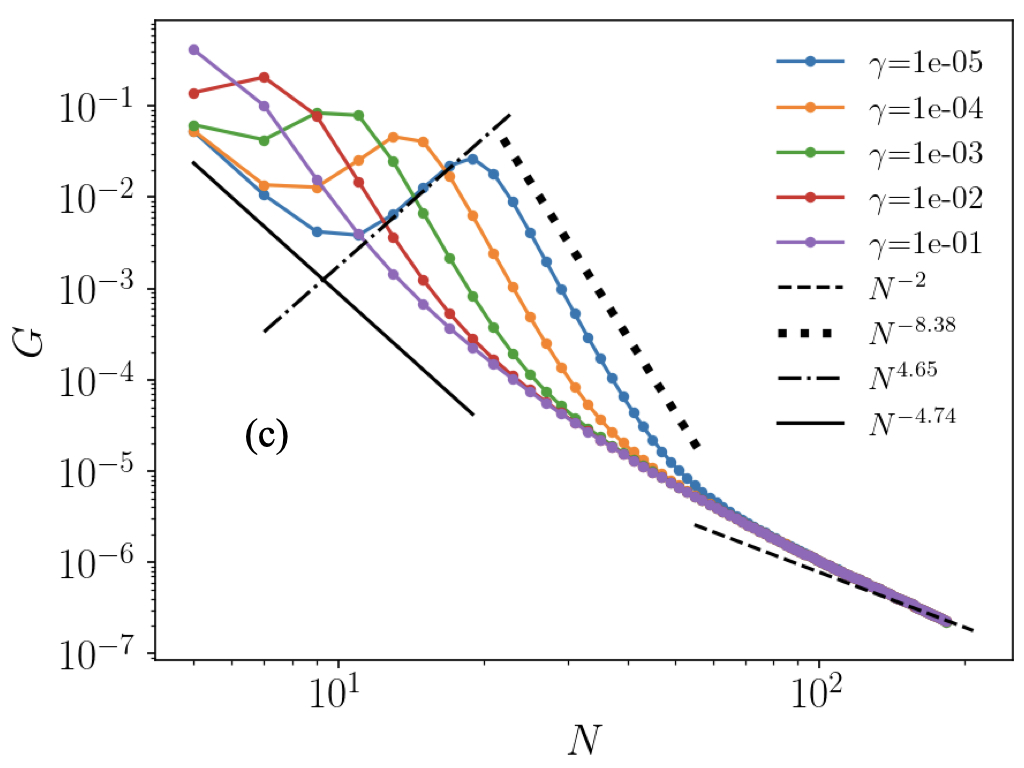}
    \caption{Scaling of exact conductance $G(\mu)$ given in Eq.~\eqref{conductance} with system size $N$ at the upper band edge corresponding to higher order exceptional points with order $p$ : (a) $p=6$ for the hopping amplitudes $J_1=1$, $J_2=2J_1/5$ and $J_3=J_2/6$. (b) $p=8$ for the hopping amplitudes $J_1=1$, $J_2=J_1/2$, $J_3=2J_2/7$ and $J_4=J_3/8$. (c) $p=10$ for the hopping amplitudes $J_1=1.0$, $J_2=4J_1/7$, $J_3=3J_2/8$, $J_4=2J_3/9$ and $J_5=J_4/10$. We take $\gamma_0=1$. The figure demonstrates the crossover from the first subdiffusive $(1/N^{\frac{p}{2}})$ to superballistic $(N^{p-2})$ to second subdiffusive $(1/N^{p})$ and to final subdiffusive $(1/N^{2})$ regime as one increases system size $N$. This behavior is summarized in Table.~\ref{EP-table}.}
    \label{high-order-EP-local}
\end{figure*}

\subsubsection{The case $n>2$}
We further extend our study for conductance scaling at the band-edge in presence of local  loss by considering the finite range model with hopping range $n=3,4$ and $5$ which possess highest order EPs of sixth, eighth and tenth order, respectively at the band-edges. In Fig.~\ref{high-order-EP-local} we demonstrate the rich transport behaviour for conductance for different orders of EPs as a function of system size by setting $\mu$ value corresponding to the upper band edge. We observe an early order of EP dependent subdiffusive scaling that goes as 
\begin{equation}
    G(\mu) \propto \frac{1}{N^{\frac{p}{2}}}, \,\, \,\,\,{\mathrm {for}} \, \, N<A_p\,\mathcal{O}\Big(\frac{1}{\gamma^{2/{(3p-4)}}}\Big),
\end{equation} 
where $p=2n$ is the highest order EP. After this initial subdiffusive regime, we once again observe interesting superballistic transport with scaling exponent that depends on the order of the EP and is given as,
\begin{align}
    G(\mu) \propto N^{p-2}.
\end{align}
 These trends can once again be understood following Eq.~\eqref{conductance-within-band-app-gr-1}. We numerically find that $|g_{1N}^r|^2$ scales as $1/N^{\frac{p}{2}}$ and $|g_{1m}^r|^2$ scales as $N^{p-2}$. As a result, we can estimate the length scale at which the superballistic scaling emerges given as $N \sim A_p\,\mathcal{O}\Big(\frac{1}{\gamma^{2/{(3p-4)}}}\Big)$ (setting $\gamma_0=1)$, where $A_p$ is a constant that depends on the order of the EP. This result indicates that the onset of superballistic transport can be controlled either by tuning $\gamma$ or by tuning the order of the EP $p$. For large $N$, eventually, in all cases, the transport becomes subdiffusive with order independent $1/N^2$ scaling. However, between the superballistic and the large $N$ subdiffusive regime, there is another interesting order dependent subdiffusive regime with scaling $1/N^{p}$ emerges. All these regimes are clearly demonstrated in Fig.~\ref{high-order-EP-local}(a), (b), and (c), for sixth, eighth, and tenth order, respectively. We summarize the order dependent dominant transport scaling for conductance and its crossover as a function of $N$ in Table.~\ref{EP-table} and we provide the estimate of the transport window here (for $p>2)$ as,
 \begin{equation}
\label{G-mu-n-ge-2}
G(\mu) \propto
    \begin{cases}
        \frac{1}{N^{\frac{p}{2}}}\quad\,\,\mathrm{for}\,\, N<A_p\,\mathcal{O}\Big(\!\frac{1}{\gamma^{2/{(3p-4)}}}\!\Big) \\ \\
        N^{p-2} \,\,\, \mathrm{for}\,\, A_p\,\mathcal{O}\Big(\!\frac{1}{\gamma^{2/{(3p-4)}}}\!\Big)\!\!<\!N\!<\!\!B_p\,\mathcal{O}\Big(\!\frac{1}{\gamma^{1/(p-1)}}\!\Big)\\ \\
        \frac{1}{N^{p}} \quad \,\,\,\mathrm{for}\,\, B_p\,\mathcal{O}\Big(\!\frac{1}{\gamma^{1/(p-1)}}\!\Big)\!\!<\!N\!<\!\!C_p\,\mathcal{O}\Big(\!\frac{1}{\gamma^{1/(p-2)}}\!\Big) \\ \\
        \frac{1}{N^{2}}\quad \,\,\,\mathrm{for}\,\, N>C_p\,\mathcal{O}\Big(\!\frac{1}{\gamma^{1/(p-2)}}\!\Big)
    \end{cases}
\end{equation}
We estimate the constants for sixth order EP as $A_6 \approx 4$, $B_6 \approx 8$ and $C_6 \approx 25$. Similarly for the eighth order EP, we estimate the constants as $A_8 \approx 5$, $B_8 \approx 10$, and $C_8 \approx 16$.

\section{Multiple non-extensive loss and role of symmetry}\label{sec-IV}
In previous section (Sec.~\ref{sec-III}), we studied the NESS transport in presence of single loss located at the middle site of the lattice. In this section, we discuss the fate of transport when more than one lossy channel but non-extensive in number is present. Before discussing this scenario we first discuss the role of symmetry of our lattice setup (Fig.~\ref{schematic-local-loss}) studied in previous section (Sec.~\ref{sec-III}).

\begin{figure}
    \centering
    \includegraphics[width=8.5cm]{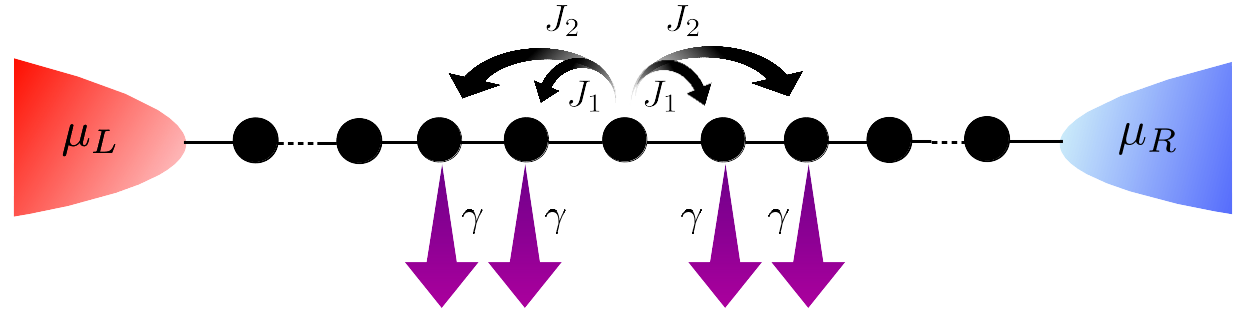}\\
    \vspace{1cm}
    \includegraphics[width=8.5cm]{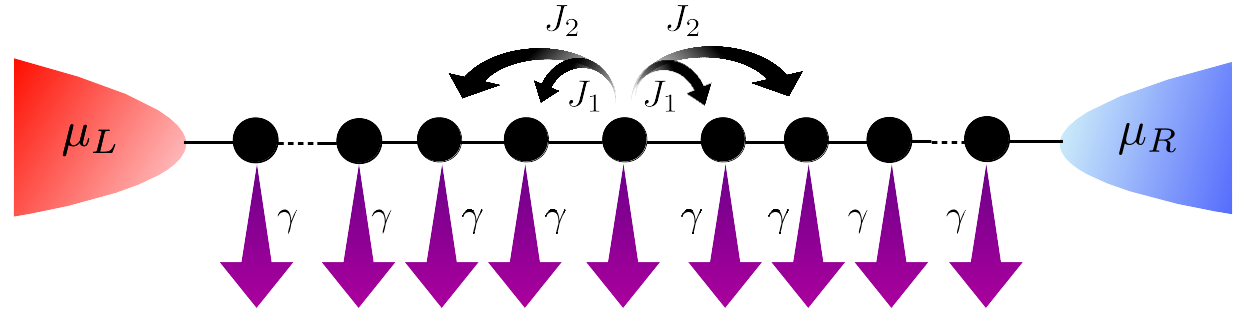}
    \caption{[Top panel]: The lattice discussed in the schematic Fig.~\ref{schematic-local-loss} is subjected to multiple but non-extensive number of local lossy channels respecting mirror symmetry about the middle site. [Bottom Panel]: The lattice is subjected to extensive number of lossy channels i.e., one at each site. The strength of the lossy channels are assumed to be the same ($\gamma$).}
    \label{schematic-extensive-loss}
\end{figure}

{\it Role of symmetry in case of single lossy channel}-- The condition $\gamma_L=\gamma_R$ and the local loss present at the middle site of the lattice gives rise to the following symmetry relation for arbitrary finite range of hopping $n$,
\begin{align}
    \mathcal{G}^{r}_{1j}(\omega)= \mathcal{G}^{r}_{N-j+1, N}(\omega). \label{reciprocity-1}
\end{align}
where recall that $\mathcal{G}^r(\omega)$ is the NEGF given in Eq.~\eqref{GF}. This symmetry relation can be easily verified numerically. Here we provide an analytical proof for $n=1$. Using Eq.~\eqref{M_element}, we can write,
\begin{align}
    &\mathcal{G}^{r}_{1j}(\omega)=-\frac{1}{J}(-1)^{j+1}\frac{\Delta_{j+1,N}(\omega)}{\Delta_{1,N}(\omega)},\\
    &\mathcal{G}^{r}_{N-j+1,N}(\omega)=-\frac{1}{J}(-1)^{1-j}\frac{\Delta_{1,N-j}(\omega)}{\Delta_{1,N}(\omega)}.
\end{align}
Using the transfer matrix $\mathbb{T}_{0}(\omega)$ given in Eq.~\eqref{T0}, we can write the following equations for $\Delta_{j+1,N}(\omega)$ and $\Delta_{1,N-j}(\omega)$ components,
\begin{widetext}
    \begin{align}
        &\begin{pmatrix}
            \Delta_{j+1,N}(\omega)\\
            \Delta_{j+2,N}(\omega)\\
        \end{pmatrix}=\Big[\mathbb{T}_{0}(\omega)\Big]^{N-j}\begin{pmatrix}
            1 \\
            \frac{i\gamma_R}{2J}
        \end{pmatrix},\quad (\mathrm{for}\,\, j+1 > m) \label{B1}\\
        &\begin{pmatrix}
            \Delta_{j+1,N}(\omega) \\
            \Delta_{j+2,N}(\omega) \\
        \end{pmatrix}=\Big[\mathbb{T}_0(\mu)\Big]^{\frac{N-1}{2}-j}\begin{pmatrix}
            -\frac{\omega}{J}-\frac{i\gamma}{2J} & \,\,\,\,-1 \\
            1  & 0 \\
        \end{pmatrix}\Big[\mathbb{T}_{0}(\mu)\Big]^{\frac{N-1}{2}}\begin{pmatrix}
            1 \\
           \frac{i\gamma_R}{2J} \\
        \end{pmatrix},\,\,\,\,\,\,(\mathrm{for}\,\,j+1\le m)\label{B2}\\
        &\begin{pmatrix}
            \Delta_{1,N-j}(\omega)\\
            \Delta_{1,N-j-1}(\omega)\\
        \end{pmatrix}=\Big[\mathbb{T}_{0}(\omega)\Big]^{N-j}\begin{pmatrix}
            1 \\
            \frac{i\gamma_L}{2J}
        \end{pmatrix},\quad (\mathrm{for}\,\, N-j < m) \label{B3}\\
        &\begin{pmatrix}
            \Delta_{1,N-j}(\omega) \\
            \Delta_{1,N-j-1}(\omega) \\
        \end{pmatrix}=\Big[\mathbb{T}_0(\mu)\Big]^{\frac{N-1}{2}-j}\begin{pmatrix}
            -\frac{\omega}{J}-\frac{i\gamma}{2J} & \,\,\,\,-1 \\
            1  & 0 \\
        \end{pmatrix}\Big[\mathbb{T}_{0}(\mu)\Big]^{\frac{N-1}{2}}\begin{pmatrix}
            1 \\
           \frac{i\gamma_L}{2J} \\
        \end{pmatrix},\,\,\,\,\,\,(\mathrm{for}\,\,N-j\ge m). \label{B4}
    \end{align}
\end{widetext}
Comparing Eq.~\eqref{B1} with Eq.~\eqref{B3} and similarly comparing Eq.~\eqref{B2} with  Eq.~\eqref{B4}, one can notice that once we set $\gamma_L=\gamma_R=\gamma_0$, 
\begin{align}
\label{symmetry-Delta}
\Delta_{j+1,N}(\omega)=\Delta_{1,N-j}(\omega),
\end{align}
which further leads to the symmetry relation for the NEGF as given in Eq.~\eqref{reciprocity-1}.
It is important to note that this symmetry relation holds only when the loss channel is located at the middle site. Moreover, it is easy to verify this relation  for arbitrary range of hopping $n$.

As a consequence of this symmetry relation, the transmission functions characterizing the loss from the left [$\mathcal{L}_L(\omega)$] and  the right [$\mathcal{L}_R(\omega)$] reservoirs become equal i.e., $\mathcal{L}_L(\omega) = \mathcal{L}_R(\omega)=\mathcal{L}(\omega)$ [see Eq.~\eqref{L-mu}]. In fact, interestingly,  under this symmetry condition the net current flowing through the system from the left to the right reservoir is insensitive to the local chemical potential at the middle site of the lattice. Thus, under this condition, the conductance in presence of localized loss at the middle site become exactly the same with that of a local gain channel [i.e., replace $c_m \to c_m^{\dagger}, c_m^{\dagger} \to c_m$ in Eq.~\eqref{density-loss}] present at the middle site of the lattice. Moreover, since the current is insensitive to the chemical potential at the middle site, if one connects a zero temperature B$\ddot{\mathrm{u}}$ttiker voltage probe \cite{Buttiker,PhysRevB.75.195110,korol2018probezt} [another particle reservoir that fixes its chemical potential such that no net current flows into the reservoir], the conductance remains the same (see Appendix \ref{appendix-B} for the proof).

It is worth pointing out that if the location of the loss channel is not at the middle site of the lattice, the symmetry relation given in Eq.~\eqref{reciprocity-1} will be violated and as a consequence the net current through the lattice can not be expressed as a difference of the Fermi functions of the two boundary reservoirs, as in Eq.~\eqref{left-right-current}.  In fact, in such a scenario, the net current $I$ is given by,
\begin{align}
    I\!=\!\int_{-\infty}^{\infty} \frac{d\omega}{2\pi} \Big[T(\omega)(f_{L}\!-f_R)+\mathcal{L}_{L}(\omega)f_{L}-\mathcal{L}_{R}(\omega)f_R\Big],\label{asymmetric-current}
\end{align}
where recall that $T(\omega)=\gamma_0^2 |\mathcal{G}^{r}_{1N}(\omega)|^{2}$, $\mathcal{L}_{L}(\omega)=\gamma \gamma_0|\mathcal{G}^{r}_{1j}(\omega)|^{2}$ and $\mathcal{L}_{R}=\gamma \gamma_{0}|\mathcal{G}^{r}_{jN}(\omega)|^{2}$ with $j$ being the lossy site. Therefore the situation when the loss is not at the middle is ill-suited for studying conductance.

{\it Multiple non-extensive lossy channels respecting symmetry}-- For every lossy channel, not located at the middle [say $j$-th site with $j<m$], if one adds an extra lossy channel, with the same loss strength, on the other side of the middle site [i.e., $(N-j+1)$-th site], then the symmetry in Eq.~\eqref{reciprocity-1} is restored. More generally, one can have multiple loss channels [see top panel of Fig.~\ref{schematic-extensive-loss}] each of which have a mirror channel about the middle site. In that case, the conductance can be expressed as 
\begin{equation}
G(\mu)=\gamma_0^{2}|\mathcal{G}^{r}_{1N}(\mu)|^{2}+\frac{\gamma\gamma_0}{2}\sum_{l} |\mathcal{G}^{r}_{1,m-l}(\mu)|^{2},
\label{conductance-non-extensive}
\end{equation}
where $l \in \pm 1, \pm 2 \cdots$ represents the location of the loss channels as viewed from the middle site. We would like to remark that in this section we assume that the number of lossy channels are non-extensive i.e., $\mathcal{O}(1)$. 

Our general observation is that when the upper band-edge corresponds to a $p$-th order EP, then in presence of at least $p/2$ (recall that $p$ is always even at the band-edge) lossy channels, placed symmetrically about the middle site, the conductance shows a cascade of superballistic regimes  with dominant system size scaling $N^{p-\alpha}$ where $\alpha$ takes the values $2, 4, \cdots p-2$ for the successive superballistic regimes starting from the first one. All of these superballistic transport regimes followed by a subdiffusive regimes with dominant system size scaling $1/N^{p-\beta}$ with $\beta$ takes the values $0, 2, 4, \cdots p-4$ for the successive subdiffusive regimes. Finally, after the $1/N^4$ subdiffusive regime, the conductance shows a ballistic $N^{0}$ scaling followed by the universal $1/N^2$ scaling in the large $N$ limit.

\begin{figure}
    \centering
    \includegraphics[width=4.2cm]{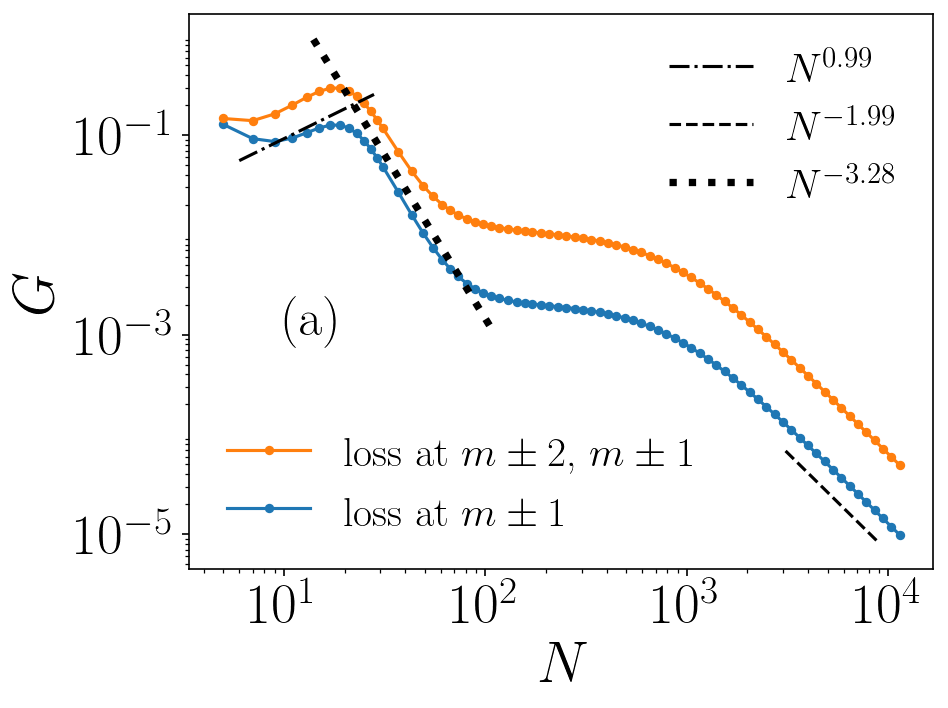}%
    \includegraphics[width=4.2cm]{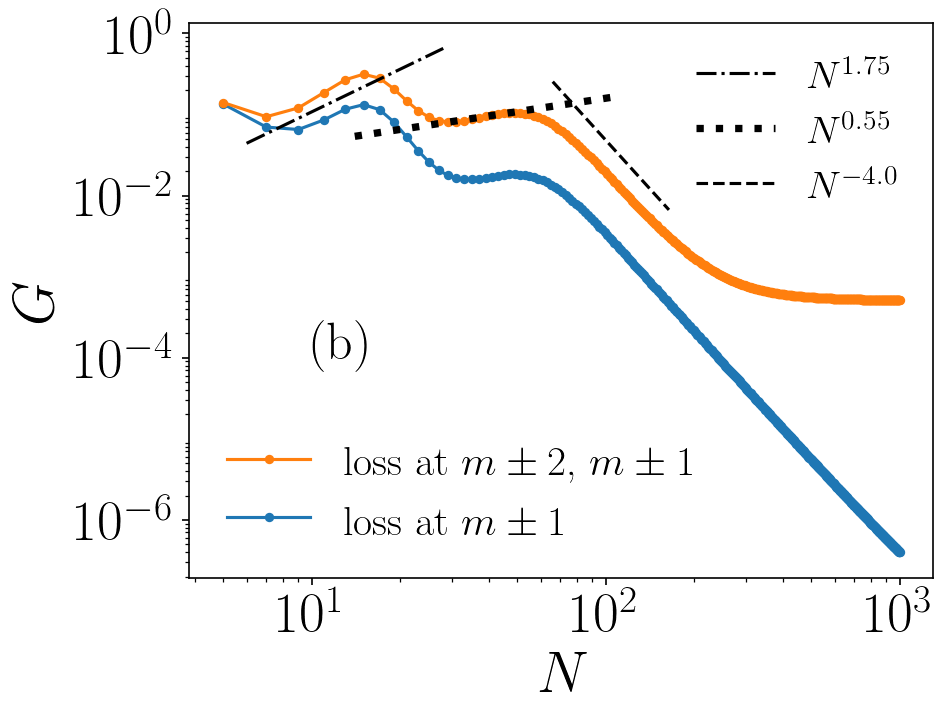}
    \caption{Scaling of exact conductance $G(\mu)$ given in Eq.~\eqref{conductance-non-extensive} with system size $N$ in presence of multiple nonextensive  loss  channels [see top panel of Fig.~\ref{schematic-extensive-loss}] for fourth and sixth order EPs. We choose two different configurations for both EPs: (i) two loss channels placed symmetrically next to the middle site $l=\pm 1$, (ii) four loss channels with two loss channels next to the middle site ($l=\pm 1)$ and the two other loss channels are located next to next to middle site ($l=\pm 2)$. (a) Fourth order EP at the upper band-edge corresponding to the chemical potential $\mu= 3 J_1/2$ for range of hopping $n=2$.   Here $\gamma_0=1, \gamma=0.005, J_1=1$, and $J_2=1/4$. The analytical arguments for the observed behaviour is presented in detail in Appendix \ref{appendix-C}. (b) Sixth order EP at the upper band-edge corresponding to the $\mu=4J_1/3$ for range of hopping $n=3$. Here the other parameters are $\gamma_0=1$, $\gamma=0.001$, $J_1=1.0$, $J_2=2/5$, $J_3=1/15.$}
    \label{multiple-nonextensive-loss}
\end{figure}

In Fig.~\ref{multiple-nonextensive-loss}, we illustrate the results for conductance scaling for range of hopping $n=2$ and $n=3$ in presence of two lossy channels ($l =\pm 1$) and four loss channels ($l =\pm 2$), always placed symmetrically about the middle site. For $n=2$ [Fig.~\ref{multiple-nonextensive-loss}(a)] when the upper band-edge has highest fourth order EP, in presence of two or four lossy channels we see that the conductance goes through a superballistic to subdiffusive $1/N^{4}$ scaling to ballistic to final subdiffusive $1/N^{2}$ regime. We provide analytical support to these observations in Appendix \ref{appendix-C}. For $n=3$ [Fig.~\ref{multiple-nonextensive-loss}(b)] i.e., for sixth order EP at the upper band-edge,  we see the appearance of the superballistic cascade when four lossy channels are present. As mentioned above, we further observe the ballistic plateau that appears after the cascade followed by the $1/N^2$ subdiffusive transport regime. 

Having observed rich conductance scalings when the losses are non-extensive, a natural next question that arises is what happens to conductance with extensive number of lossy channels which we study next. 

\section{Extensive loss}
\label{sec-5}
In this section, we study the conductance scaling for chemical potential located at the band edges when extensive number of lossy channels are present in the lattice. In fact, we consider a situation in which each lattice site suffers particle loss with loss rate $\gamma$ [see bottom panel of Fig.~\ref{schematic-extensive-loss}]. The corresponding Lindblad master equation takes the form 
\begin{equation}
    \frac{d\rho}{dt}=-i\big[H,\rho\big]+ \gamma \sum_{i=1}^{N}  \Big[c_{i}\rho c_{i}^{\dagger}-\frac{1}{2}\{c_{i}^{\dagger} c_{i},\rho\}\Big].
    \label{density-loss-extensive}
\end{equation}

\begin{figure}[h]
    \centering
    \includegraphics[width=7cm]{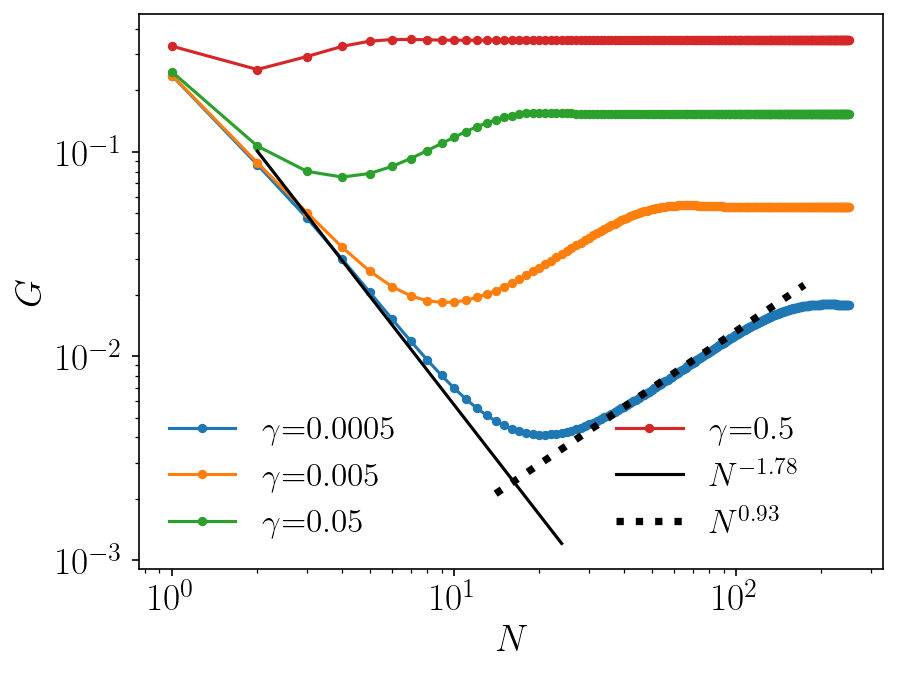}
    \caption{Exact conductance in Eq.~\eqref{conductance-extensive-loss} vs system size scaling at the upper band-edge of the nearest neighbour hopping model i.e., $n=1$ in presence of extensive number of local loss [see bottom panel of Fig.~\ref{schematic-extensive-loss}] in the system. Here the band-edge possess a second order EP. Here $\gamma_0=J=1$.}
    \label{2nd-order-extensive}
\end{figure}

In presence of such extensive loss with same loss strength $\gamma$, the symmetry relation involving the Green's function components, given in Eq.~\eqref{reciprocity-1}, is always preserved as long as $\gamma_L=\gamma_R=\gamma_0$. In this case, the expression of the net current flowing through the system from the left to the right reservoir (i.e., in the direction of the chemical bias) is given as,
\begin{align}
    I=\int_{-\infty}^{+\infty} \frac{d\omega}{2\pi} \Big[T(\omega)+\frac{1}{2}\sum_{i=1}^{N} \mathcal{L}_{i}(\omega)\Big]\big[f_L(\omega)-f_{R}(\omega)\big]. \label{current-extensive-loss}
\end{align}
Recall that, $T(\omega)=\gamma_0^{2}\,|\mathcal{G}^{r}_{1N}(\omega)|^{2}$ and $\mathcal{L}_{i}(\omega)=\gamma\gamma_0\, |\mathcal{G}^{r}_{1i}(\omega)|^{2}$ where the NEGF $\mathcal{G}^{r}(\omega)$ now becomes,
\begin{align}
    \mathcal{G}^{r}(\omega)=\Big[\omega \mathbb{I}_{N}-h_{S}-\Sigma_L-\Sigma_R+i\Gamma\Big]^{-1},
    \label{Gr-extensive-loss}
\end{align} 
with the only change occuring in the $N\times N$ $\Gamma$ matrix whose elements are now given by  $\Gamma_{ij}=\gamma\,\delta_{ij}/2$. As a result, the conductance  at zero temperature and in the linear response limit reduces to,
\begin{align}
    G(\mu)=\gamma_0^{2}\,|\mathcal{G}^{r}_{1N}(\mu)|^{2}+\frac{\gamma\gamma_0}{2}\sum_{i=1}^{N}|\mathcal{G}^{r}_{1i}(\mu)|^{2}. \label{conductance-extensive-loss}
\end{align}
We now first study the conductance scaling with system size in presence of such extensive number of lossy channels for the nearest neighbour hopping model i.e., $n=1$. This is illustrated in Fig.~\ref{2nd-order-extensive} where we observe a crossover from the initial $1/N^2$ subdiffusive transport regime to superballistic transport regime and then finally to a ballistic regime. It is interesting to note that, for such extensive loss case, the intermediate superballistic regime appears even for $n=1$. Recall that for local loss [see Sec.~\ref{sec-III}] for $n=1$ case, the intermediate regime did not have any superballistic behaviour [see Fig.~\ref{2nd-order-local}] but rather at most a ballistic plateau (this is also true even for non-extensive number of lossy channels for $n=1$). We now show analytically how these different transport regimes emerge for such extensive loss setup for $n=1$. 

We first explain the origin of initial subdiffusive $1/N^2$ scaling and the onset of superballistic transport.
Since we observe in Fig.~\ref{2nd-order-extensive} that the superballistic transport regime appears for small $\gamma \ll 1$, we write the conductance expression, given in Eq.~\eqref{conductance-extensive-loss}, upto the first non-zero order of $\gamma$ to obtain,
\begin{align}
G(\mu)=\gamma_{0}^{2}|\mathcal{G}^{r(0)}_{1N}(\mu)|^{2}+\frac{\gamma\gamma_0}{2}\sum_{i=1}^{N}|\mathcal{G}^{r(0)}_{1i}(\mu)|^{2} + \mathcal{O}(\gamma^2), \label{superballistic-extensive}
\end{align}
where $\mathcal{G}^{r(0)}(\mu)$ is the Green's function without the loss self energy term i.e.,
\begin{align}
    \mathcal{G}^{r(0)}(\mu)=\Big[\mu \mathbb{I}_{N}-h_{S} + i \Gamma_0\Big]^{-1}. \label{bareGF}
\end{align}
where recall that $\Gamma_0|_{11} = \Gamma_0|_{NN} = \gamma_0/2$ and zero otherwise.
The scaling of the elements of $\mathcal{G}^{r(0)}(\mu)$ in Eq.~\eqref{2nd-order-extensive} at the band-edge can be obtained in a similar way following Eqs.~\eqref{GF_element}-\eqref{T5} by substituting the loss strength $\gamma=0$. Recall that the elements of the Green's function such as $\mathcal{G}^{r(0)}_{1j}(\mu)$ depends on the determinants $\Delta_{1,N}(\mu)$, $\Delta_{j+1,N}(\mu)$ which are related to the transfer matrix [see Eq.~\eqref{GF_element}]. Now at the band-edge which hosts a second-order EP, $\Delta_{1,N}(\mu)$ and $\Delta_{j+1,N}(\mu)$ can be obtained from Eq.~\eqref{delta-exact-1} by substituting $\gamma=0$ and given as (setting $\gamma_0=1$),

\begin{align}
    &\Delta_{1,N}(\mu)=\frac{1}{32}\Big[2i(N+1)-(N-1)\Big]\Big[8i+16\Big], \\
    &\Delta_{j+1,N}=(-1)^{N-j}\Bigg[\frac{N-j+1}{16}-i\,\frac{j}{2}+ i \,\frac{N}{2}\Bigg].
\end{align}
Thus the scaling of the Green's function appearing in Eq.~\eqref{superballistic-extensive} at the band-edge becomes,
\begin{align}
\label{Gr1N0-extensive}
    &|\mathcal{G}^{r(0)}_{1N}(\mu)|^{2}=\frac{1}{|\Delta_{1,N}(\mu)|^2}\propto \frac{1}{N^{2}},\\
    &|\mathcal{G}^{r(0)}_{1i}(\mu)|^{2}=\frac{|\Delta_{i+1,N}|^{2}}{|\Delta_{1,N}(\mu)|^2}\propto N^0.
    \label{Gr1i-extensive}
\end{align}
Thus following Eq.~\eqref{superballistic-extensive} and using Eqs.~\eqref{Gr1N0-extensive} and \eqref{Gr1i-extensive}, we observe that the conductance $G(\mu)$ follows the subdiffusive scaling,
\begin{align}
    G(\mu)\propto \frac{1}{N^{2}}, \, \, {\mathrm {for}}\,\,\,  N< A_2\,\mathcal{O}\Big(\frac{1}{\gamma^{1/3}}\Big).
\end{align}
where in the extensive loss case the constant $A_2 \approx 2 $. On the other hand, the conductance follows superballistic scaling, 
\begin{align}
    G(\mu)\propto \gamma\sum_{i=1}^{N} N^0 =\gamma N, \,\,\,   {\mathrm {for}} \,\,\, N> A_2\,\mathcal{O}\Big(\frac{1}{\gamma^{1/3}}\Big).
    \label{conductance-extensive-superballistic}
\end{align}
This analysis therefore explains the crossover from subdiffusive to superballistic transport and provides an estimate of the crossover length scale.

We next discuss the crossover from the superballistic to the ballistic regime and estimate the corresponding crossover length scale. It turns out that at relative larger values of system size $N$, the perturbative expansion given in Eq.~\eqref{superballistic-extensive} is ill-suited to estimate the final crossover from superballistic to ballistic scaling regime. Therefore one must resort to computing the nonperturbative NEGF ${\cal G}^r(\omega)$ following Eq.~\eqref{Gr-extensive-loss}.
\begin{figure*}
    \centering
    \includegraphics[width=6.1cm]{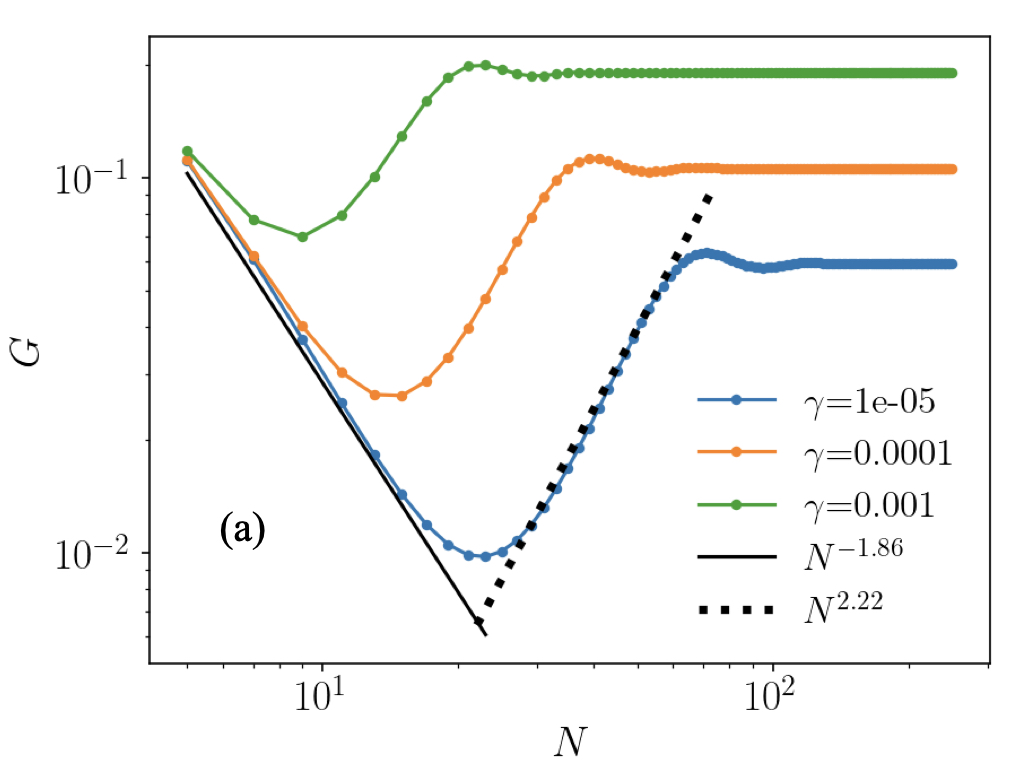}%
    \includegraphics[width=6.1cm]{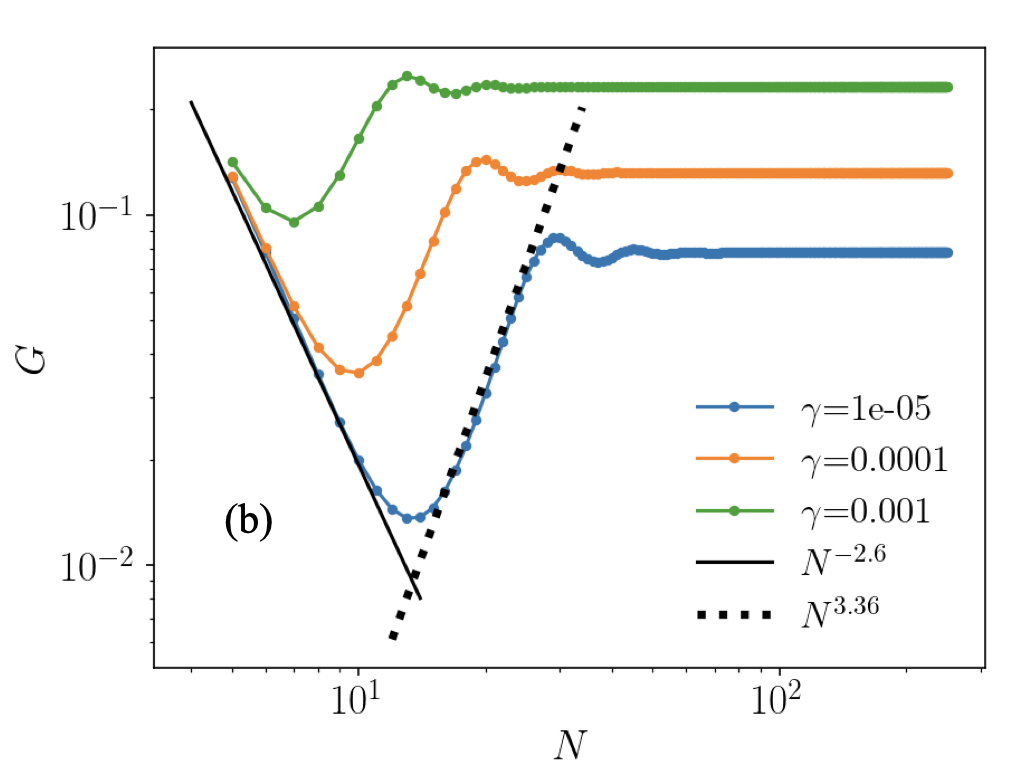}%
    \includegraphics[width=6.1cm]{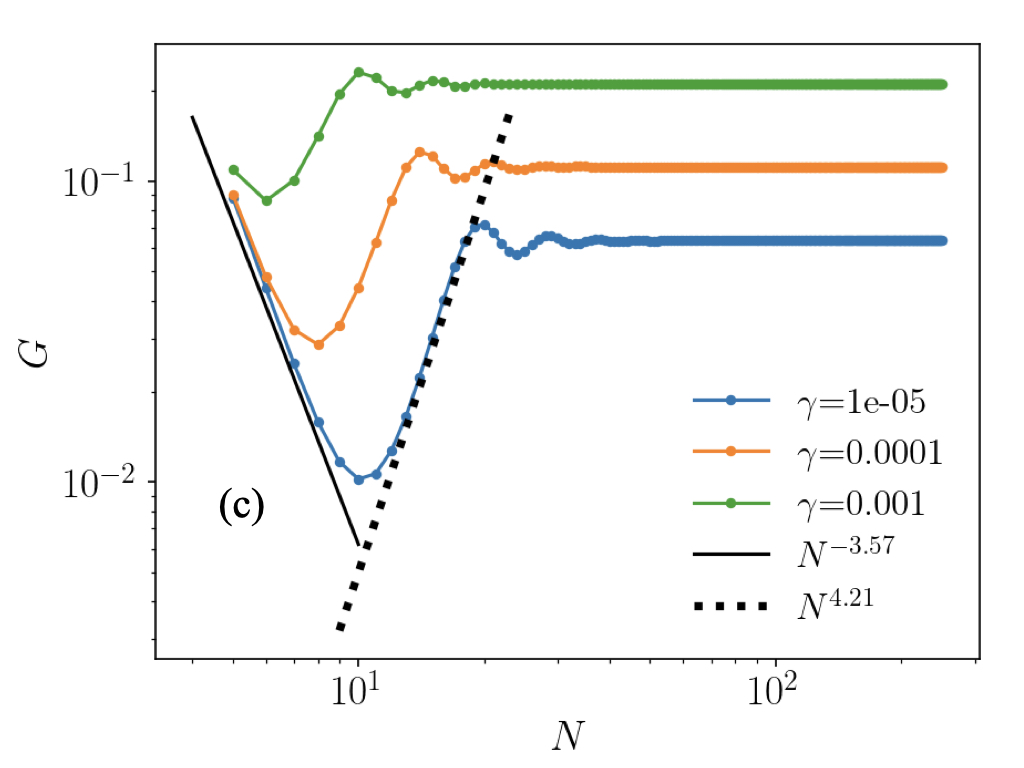}
\caption{Exact conductance in Eq.~\eqref{conductance-extensive-loss} with system size scaling in presence of extensive number of lossy channels [see bottom panel of Fig.~\ref{schematic-extensive-loss}] at the upper band-edge (corresponding to $k_{b}=\pi$) for higher order EPs with order $p$ - (a) $p=4$ ($n=2$) for the hopping amplitudes $J_1=1.0$ and $J_2=J_1/4$. (b) $p=6$ ($n=3$)for the hopping amplitudes $J_1=1$, $J_2=2J_1/5$ and $J_3=J_2/3$. (c) $p=8$ ($n=4$) for the hopping amplitudes $J_1=1$, $J_2=J_1/2$, $J_3=2J_2/7$ and $J_4=J_3/8$. We set $\gamma_0=1$.}
    \label{high-order-EP-extensive}
\end{figure*}
We will use a similar approach as in Sec.~\ref{sec:2} [see Eq.~\eqref{GF_element}-Eq.~\eqref{M_element}]. To compute the NEGF ${\cal G}^r(\omega)$, we write,
    \begin{align}
    &\begin{pmatrix}
            \Delta_{1,N}(\mu) \\
            \Delta_{2,N}(\mu) \\
        \end{pmatrix}=\begin{pmatrix}
            1 & -\frac{i\gamma_0}{2} \\
            0 & 1 \\ 
        \end{pmatrix}\Big[\mathbb{T}(\mu)\Big]^N\begin{pmatrix}
            1 \\
            \frac{i\gamma_0}{2}
        \end{pmatrix}, \label{TT1}
\end{align}
\begin{align}
        &\begin{pmatrix}
            \Delta_{1,l-1}(\mu) \\
            \Delta_{2,l-1}(\mu) \\
        \end{pmatrix}=\begin{pmatrix}
            1 & -\frac{i\gamma_0}{2} \\
            0 & 1 \\ 
        \end{pmatrix}\Big[\mathbb{T}(\mu)\Big]^{l-1}\begin{pmatrix}
            1 \\
           0 \\
        \end{pmatrix},\label{TT2}
\end{align}
\begin{align}
        &\begin{pmatrix}
            \Delta_{l,N}(\mu) \\
            \Delta_{l+1,N}(\mu) \\
        \end{pmatrix}=\Big[\mathbb{T}(\mu)\Big]^{N-l+1}\begin{pmatrix}
            1 \\
           \frac{i\gamma_0}{2} \\
        \end{pmatrix}.\label{TT4}
\end{align}
where $\mathbb{T}(\mu)$ is a $2 \times 2$ matrix given by,
\begin{equation}
\mathbb{T}(\mu)=\begin{pmatrix} -\frac{\mu}{J}+\frac{i\gamma}{2} & -1 \\ 1 & 0 \\
\end{pmatrix}.
\label{T-extensive-loss}
\end{equation}
Note that because of the presence of imaginary term $i\gamma/2$ in the first diagonal term of the matrix $\mathbb{T}(\mu)$ in Eq.~\eqref{T-extensive-loss}, it is now diagonalizable even at the band edges ($|\mu|=2J)$. Using Eqs.~\eqref{TT1}-\eqref{TT4}, we obtain the Green's function elements $|\mathcal{G}^{r}_{ij}(\mu)|^{2}$ defined in Eq.~\eqref{Gr-extensive-loss} for arbitrary finite value of $\mu$ as,
\begin{align}
    |\mathcal{G}^{r}_{ij}(\mu)|^{2} \propto \frac{1}{\big|\lambda_{+}(\mu,\gamma)\big|^{2|i-j|}},
\end{align}
where $\lambda_{+}(\mu,\gamma)$ is the eigenvalue of the $2\times 2$ matrix $\mathbb{T}(\mu)$ in Eq.~\eqref{T-extensive-loss} with the largest real part and more precisely $|\lambda_{+}(\mu,\gamma)|>1$. Thus for any value of $\mu$, the first term for conductance in Eq.~\eqref{conductance-extensive-loss}
\begin{equation}
|\mathcal{G}^{r}_{1N}(\mu)|^{2} \propto  \frac{1}{\big|\lambda_{+}(\mu,\gamma)\big|^{2N}}
\end{equation}
decays exponentially with system size i.e.,
\begin{equation}
    |\mathcal{G}^{r}_{1N}(\mu)|^{2} \propto e^{-2N \,{\rm ln}|\lambda_{+}(\mu,\gamma)|}\,. \label{localization-gr}
\end{equation}
Therefor the localization length from Eq.~\eqref{localization-gr} can be estimated to be 
\begin{equation}
    \xi \sim \frac{1}{2 \,{\rm ln}|\lambda_{+}(\mu,\gamma)|}. \label{localization-length}
\end{equation}
The second term in the conductance expression in Eq.~\eqref{conductance-extensive-loss} is
\begin{equation}
\sum_{i=1}^{N}|\mathcal{G}^{r}_{1i}(\mu)|^{2} \propto \sum_{i=1}^{N} \frac{1}{\big|\lambda_{+}(\mu,\gamma)\big|^{2|i-1|}}. \label{gr-sum-extensive}
\end{equation}
Eq.~\eqref{gr-sum-extensive} in $N\to \infty$ limit, corresponds to an infinite geometric series, eventually yielding a number which is $\mathcal{O}(N^0)$. Hence in presence of extensive number of lossy channels, the conductance eventually become ballistic in the thermodynamic limit. The crossover length scale to this ballistic regime from the superballistic regime can be estimated as follows. It turns out that the localization length given in Eq.~\eqref{localization-length} is $\mathcal{O}(1/\sqrt{\gamma})$ for small $\gamma$. This naturally helps us estimate the crossover scale between superballistic to ballistic transport. Finally our results for extensive loss can be summarized as
\begin{equation}
\label{G-n2_extensive}
   G(\mu)  \propto 
\begin{cases} 
\frac{1}{N^2} \quad \quad {\mathrm {for}} \, \, \,\, N< A_2\,\mathcal{O}\Big(\frac{1}{\gamma^{1/3}}\Big)  \\
 N    \quad \quad {\mathrm {for}} \, \, \, \,  A_2\,\mathcal{O}\Big(\frac{1}{\gamma^{1/3}}\Big)  < N< B_2\,\mathcal{O}(\frac{1}{\sqrt{\gamma}})  \\
N^0  \quad \quad {\mathrm {for}} \, \,\, \,  N> B_2\,\mathcal{O}(\frac{1}{\sqrt{\gamma}})
\end{cases} 
\end{equation}
where the constants are given by $A_2 \approx 2 $ and $B_2 \approx 1$. Recall that the analytical arguments presented above hold for $n=1$ where the highest order EP is $p=2$. It turns out that even for the higher order EPs, we observe a similar superballistic scaling of conductance followed by the ballistic scaling which we have presented in Fig.~\ref{high-order-EP-extensive}.

\section{Conclusions and Outlook}
\label{sec:conc}
In summary, in this work we have investigated steady state quantum transport properties in finite range lattices that are subjected to particle loss. In particular, we classify different transport regimes by studying system size scaling of conductance in presence of single, multiple non-extensive, and extensive number of lossy channels. We reveal that for single and non-extensive lossy channels multiple anomalous transport regimes can appear at the band-edges due to interesting interplay between the order of the exceptional points of the transfer matrix of the lattice (without the loss) and the lossy channels [see Figs.~\ref{2nd-order-local}, \ref{4th_order_local}, and \ref{high-order-EP-local}, \ref{multiple-nonextensive-loss}]. The appearance of these transport regimes can be analytically explained by using the transfer matrix technique. This approach further provides an estimate of the window of these different transport regimes and thereby enable controlling the window by tuning the loss strength $\gamma$. Some of these results are summarized in Table.~\ref{EP-table}. The conductance scaling other than at band edges remains robust even in presence of non-extensive lossy channels. This scenario however completely changes in presence of extensive number of lossy channels. Independent of the value of the chemical potential the conductance become ballistic in the thermodynamic limit [see Fig.~\ref{2nd-order-extensive}]. At the band-edges, however, the ballistic regime appears after crossing over from the superballistic regime [see Fig.~\ref{high-order-EP-extensive}]. 

Future work will be directed towards understanding the impact of lossy channels for lattice systems that contain pairing terms in the Hamiltonian (for example, Kitaev chain) \cite{PhysRevB.102.224512}. The interplay between such pairing term and the lossy channels could lead to interesting physics \cite{marco_schiro_loss}. It will also be interesting to study the role of quantum many-body interactions in presence of particle loss. In particular,  one can implement state-of-the art techniques such as Matrix Product Operator (MPO) to explore transport in such lossy systems \cite{PhysRevLett.106.220601,PhysRevA.109.L050201,PhysRevLett.93.207204}.

\section{Acknowledgements}
B. K. A. acknowledges CRG Grant No. CRG/2023/003377 from SERB, Government of India. B. K. A. would like to acknowledge funding from the National Mission on Interdisciplinary  Cyber-Physical  Systems (NM-ICPS)  of the Department of Science and Technology,  Govt.~of  India through the I-HUB  Quantum  Technology  Foundation, Pune, India. K.G. would like to acknowledge the Prime Minister's Research Fellowship (ID- 0703043), Government of India for funding. 
M.K. would like to acknowledge support from the project 6004-1 of the Indo-French Centre for the Promotion of Advanced Research (IFCPAR).  M. K. thanks the VAJRA faculty scheme (No.~VJR/2019/000079) from the Science and Engineering Research Board (SERB), Department of Science and Technology, Government of India. M. K. acknowledges support of the Department of Atomic Energy, Government of India, under Project No. RTI4001. M. K. thanks the hospitality of Department of Physics, IISER, Pune. M. K. thanks the hospitality of the Department of Physics, University of Crete (UOC) and the Institute of Electronic Structure and Laser (IESL) - FORTH, at Heraklion, Greece. We thank Madhumita Saha for providing valuable insights.  B. K. A and K. G. thanks the hospitality of International Centre of Theoretical Sciences (ICTS), Bangalore, India.

\appendix
\renewcommand{\thefigure}{B\arabic{figure}}
\setcounter{figure}{0}
\section{Analytical proof of different transport regimes for $n=2$, in presence of fourth order exceptional point at the band-edge for the case of single loss} 
\label{appendix-A}
In this appendix, we demonstrate analytically how the different transport regimes shown in Fig.~\ref{4th_order_local} of the main text emerge. The  range of hopping is chosen to be $n=2$ which therefore hosts the highest fourth order exceptional point. The chemical potential $\mu$ is located at the band-edge. We consider the case with single loss present at the middle site of the lattice (recall schematic Fig.~\ref{schematic-local-loss} of the main text). 
The transfer matrix corresponding to the lattice setup is given by,
\begin{align}
    \mathbb{T}_{0}(\mu)=\begin{pmatrix}
        -\frac{J_1}{J_2} & -\frac{\mu}{J_2} & -\frac{J_1}{J_2} & -1 \\
        1 & 0 & 0 & 0\\
        0 & 1 & 0 & 0\\
        0 & 0 & 1 & 0\\
    \end{pmatrix}. \label{4th-order-TM1}
\end{align}
This transfer matrix $\mathbb{T}_0(\mu)$ in Eq.~\eqref{4th-order-TM1} hosts the highest fourth-order exceptional point under the condition $J_1=4J_2$, and $\mu=6 J_2$ that corresponds to the upper band edge $k_b=\pi$ of the lattice. For these parameters, $\mathbb{T}_{0}(\mu)$ is not diagonalizable. To obtain the system size dependence we write the exact formula for the conductance 
\begin{align}
    G(\mu)=\Big[\gamma_{0}^{2} \,|\mathcal{G}^{r}_{1N}(\mu)|^{2}+\frac{\gamma\gamma_0}{2}\,|\mathcal{G}^{r}_{1m}(\mu)|^{2}\Big], \label{conductance-within-band-app}
\end{align}
where recall that $\mathcal{G}^{r}(\mu)$ is  given in Eq.~\eqref{GF} and it is the exact NEGF which is defined in presence of both the boundary reservoirs and the loss channel. The appearance of different transport regimes and the corresponding system size scaling in Fig.~\ref{4th_order_local}
can be explained by obtaining $\mathcal{G}^{r}(\mu)$ only in the presence of loss channel
and ignoring the self energy due to the boundary reservoirs. This is done for simplifying the calculation for Green's function components and the final scaling results for conductance remain the same. We denote this Green's function as $g^r(\mu)$. Now the conductance expression in Eq.~\eqref{conductance-within-band-app} becomes 
\begin{align}
    G(\mu)=\Big[\gamma_{0}^{2} \,|{g}^{r}_{1N}(\mu)|^{2}+\frac{\gamma\gamma_0}{2}\,|{g}^{r}_{1m}(\mu)|^{2}\Big] 
    \label{conductance-within-band-app-gr}
\end{align}
and is correct upto $\mathcal{O}(\gamma_0)$. We write $g^{r}(\mu)$ as 
\begin{align}
    g^{r}(\mu)=\frac{1}{J_2} \big[M(\mu)\big]^{-1}\,,\,\,\, M(\mu)=\frac{1}{J_2}\big[\mu\mathbb{I}_{N}-h_{S}+i\Gamma\big] \label{Mmu},
\end{align}
where recall that $h_S$ is the $N \times N$ single particle hamiltonian matrix for lattice with range of hopping $n=2$ and therefore is not a tridiagonal matrix. Also $\Gamma_{lj} = \gamma\,\delta_{lm}\,\delta_{jm}/2$ with $m$ being the middle site of the lattice.
We need to compute different components of the Green's function $g^{r}(\mu)$ to calculate the conductance. For that we now need to obtain the inverse of the Toeplitz matrix $M(\mu)$ \cite{LAVIS1997137,grenander1958toeplitz,BINI198399}. This can be obtained by using the transfer matrix $\mathbb{T}_{0}(\mu)$ in Eq.~\eqref{4th-order-TM1} \cite{hypersurface_madhumita}. We first use the relation $ M(\mu) M(\mu)^{-1} =\mathbb{I}_{N}$ and write,
\begin{align}
  \sum_{k=1}^{N} \langle i|M(\mu) |k\rangle\, \langle k|M(\mu)^{-1} |j\rangle = \delta_{ij}, \quad 1 \leq i, j \leq N. \label{id-rel}
\end{align}
The elements of the matrix $M(\mu)$ given in Eq.~\eqref{Mmu} is,
\begin{eqnarray}
      \langle i|M(\mu)|k\rangle &=& \frac{\mu}{J_2}\delta_{ik}+\frac{J_1}{J_2}(\delta_{i(k+1)}+\delta_{i(k-1)})\nonumber \\
      &+& (\delta_{i(k+2)}+\delta_{i(k-2)}) 
      +\frac{i\gamma}{2J_2}\delta_{im}\delta_{km}. \label{Mmu-elements}
\end{eqnarray}
We now define a $4\times 1$ dimensional vector $V_{i}(j)$ as,
\begin{align}
    V_{i}(j)=\begin{pmatrix}
        \langle i-1 | M(\mu)^{-1} | j\rangle \\
        \langle i | M(\mu)^{-1} | j\rangle \\
        \langle i+1 | M(\mu)^{-1} | j\rangle \\
        \langle i+2 | M(\mu)^{-1} | j\rangle \\
    \end{pmatrix}. \label{V-vector}
\end{align}
\begin{widetext}
Note that as the system size is $N$, the elements of the last [first] two rows of the vector $V_{N}(j)$ [$V_{0}(j)$] are $0$, i.e., 
\begin{align}
    V_{N}(j)=\begin{pmatrix}
        \langle N-1 | M(\mu)^{-1} | j\rangle \\
        \langle N | M(\mu)^{-1} | j\rangle \\
        0 \\
        0 \\
    \end{pmatrix},\quad V_{0}(j)=\begin{pmatrix}
    0\\
    0\\
        \langle 1 | M(\mu)^{-1} | j\rangle \\
        \langle 2 | M(\mu)^{-1} | j\rangle \\
    \end{pmatrix}. \label{VN-V0}
\end{align}
Using the elements of $M(\mu)$ given in Eq.~\eqref{Mmu-elements} and the vector $V_{i}(j)$ defined in Eq.~\eqref{V-vector}, we write the Eq.~\eqref{id-rel} in terms of transfer matrix $\mathbb{T}_{0}(\mu)$, given in Eq.~\eqref{4th-order-TM1} as 
\begin{align}
    &\mathbb{T}_{0}(\mu)\,V_{i}(j)=V_{i-1}(j)-\delta_{ij}\mathbb{I}_{4}|1\rangle \,\,\quad \mathrm{for}\,\, i\neq m, \label{TM-rel-other}\\
    &\Tilde{\mathbb{T}}(\mu)\,V_{i}(j)=V_{i-1}(j)-\delta_{ij}\mathbb{I}_{4}|1\rangle \,\,\quad \mathrm{for}\,\, i= m, \label{TM-rel-middle}
\end{align}
where 
\begin{equation}
\label{T-tilde}
\Tilde{\mathbb{T}}(\mu)=\Big[\mathbb{T}_{0}(\mu)-i \frac{\gamma}{2J_2}R\Big]  
\end{equation}
with $R$ being a $4\times 4$ matrix with elements $R_{ij}=\delta_{1i}\,\delta_{2j}$. Note that we need the matrix elements $\langle 1|M(\mu)^{-1}|N \rangle$ and $\langle 1|M(\mu)^{-1}|m \rangle$. To obtain this, we first use Eq.~\eqref{TM-rel-other} and Eq.~\eqref{TM-rel-middle} to write the form of $V_{N}(N)$ and $V_{N}(m)$ in terms of $V_{0}(N)$ and $V_{0}(m)$ as,
\begin{align}
    &V_{N}(N)=\big[\mathbb{T}_{0}(\mu)\big]^{-\frac{N-1}{2}} \,\Tilde{\mathbb{T}}(\mu)^{-1}\big[\mathbb{T}_{0}(\mu)\big]^{-\frac{N-1}{2}} V_{0}(N) - \big[\mathbb{T}_{0}(\mu)\big]^{-1} |1\rangle, \label{TM-rel-other1} \\
    &V_{N}(m)=\big[\mathbb{T}_{0}(\mu)\big]^{-\frac{N-1}{2}} \,\Tilde{\mathbb{T}}(\mu)^{-1}\big[\mathbb{T}_{0}(\mu)\big]^{-\frac{N-1}{2}} V_{0}(m) - \big[\mathbb{T}_{0}(\mu)\big]^{-\frac{N-1}{2}}\Tilde{\mathbb{T}}(\mu)^{-1} |1\rangle. \label{TM-rel-middle1}
\end{align}
Using the fact that $V_{N}(j)$ has zeros in the last two rows for any $j$, we equate the left and the right hand sides of Eqs.~\eqref{TM-rel-other1} and \eqref{TM-rel-middle1} for the last two rows and obtain the following relations 
\begin{align}
    &\sum_{l=1,2} \langle s+l | A | 2+l\rangle \langle l|M(\mu)^{-1}|N\rangle - \langle s+l| \mathbb{T}_{0}(\mu)^{-1}|1\rangle =0,\,\,\quad \mathrm{where}\,\, s=1,2\label{sum-2}\\
    &\sum_{l=1,2} \langle s+l | A | 2+l\rangle \langle l|M(\mu)^{-1}|m\rangle - \langle s+l| \mathbb{T}_{0}(\mu)^{-\frac{N-1}{2}}\Tilde{\mathbb{T}}(\mu)^{-1} |1\rangle =0 \,\,\quad \mathrm{where}\,\, s=1,2.\label{sum-1} 
\end{align}
Here the matrix $A$ is given as
\begin{equation}
A=\big[\mathbb{T}_{0}(\mu)\big]^{-\frac{N-1}{2}} \,\,\Tilde{\mathbb{T}}(\mu)^{-1}\,\,\big[\mathbb{T}_{0}(\mu)\big]^{-\frac{N-1}{2}}.
\label{A-matrix}
\end{equation}
Now from Eq.~\eqref{sum-2}, we can write the following matrix equation,
\begin{align}
\label{procedure-A}
    &\begin{pmatrix}
        A_{33} & A_{34} \\
        A_{43} & A_{44} \\
    \end{pmatrix}\begin{pmatrix}
        \langle 1|M(\mu)^{-1}|N\rangle \\
        \langle 2|M(\mu)^{-1}|N\rangle \\
    \end{pmatrix}=\begin{pmatrix}
        \langle 3| \mathbb{T}_{0}(\mu)^{-1}|1\rangle \\
        \langle 4| \mathbb{T}_{0}(\mu)^{-1}|1\rangle \\
    \end{pmatrix}
 \end{align}
which can alternatively be expressed as  
    \begin{align}
     \begin{pmatrix}
        \langle 1| M(\mu)^{-1}|N\rangle \\
        \langle 2| M(\mu)^{-1}|N\rangle \\
    \end{pmatrix}=\begin{pmatrix}
        A_{33} & A_{34} \\
        A_{43} & A_{44} \\
    \end{pmatrix}^{-1}
    \begin{pmatrix}
        \langle 3|\mathbb{T}_{0}(\mu)^{-1}|1\rangle \\
        \langle 4|\mathbb{T}_{0}(\mu)^{-1}|1\rangle\\
    \end{pmatrix}.
\end{align}
Finally we obtain, 
\begin{align}
    \langle 1| M(\mu)^{-1}|N\rangle=\frac{A_{44} \langle 3|\mathbb{T}_{0}(\mu)^{-1}|1\rangle\,-\,A_{34}\langle 4|\mathbb{T}_{0}(\mu)^{-1}|1\rangle}{A_{33}A_{44}-A_{34}A_{43}}.
    \label{1MN}
\end{align}
Similarly the component $\langle 1| M(\mu)^{-1}|m\rangle$ can be obtained from Eq.~\eqref{sum-1} and is given as,
\begin{align}
    \langle 1| M(\mu)^{-1}|m\rangle=\frac{A_{44} \langle 3|\mathbb{T}_{0}(\mu)^{-\frac{N-1}{2}}\Tilde{\mathbb{T}}(\mu)^{-1} |1\rangle\,-\,A_{34}\langle 4|\mathbb{T}_{0}(\mu)^{-\frac{N-1}{2}}\Tilde{\mathbb{T}}(\mu)^{-1} |1\rangle}{A_{33}A_{44}-A_{34}A_{43}}.
    \label{M1m}
\end{align}
Now the task is to compute the elements of $A$ following Eq.~\eqref{A-matrix}. We therefore need to calculate the matrix $\mathbb{T}_{0}(\mu)^{-\frac{N-1}{2}}$.
To obtain $\langle i|\mathbb{T}_{0}(\mu)^{-b}|j\rangle$ for any positive integer $b$ when $\mathbb{T}_{0}(\mu)$ is non-diagonalizable (i.e., when $\mu=6J_2$ corresponds to the upper band-edge. Henceforth we set $J_1=1$.),  we use the Jordan normal form,
\begin{align}
\label{jordan-normal-j}
 J=S \,\mathbb{T}_{0}(\mu)\, S^{-1}= \begin{pmatrix}
       -1 & 1 & 0 & 0\\
        0 & -1 & 1 & 0\\
        0 & 0 & -1 & 1\\
        0 & 0 & 0 & -1\\
 \end{pmatrix},\,\,\,\,\,\mathrm{where}\,\,\, S=\begin{pmatrix}
        0 & 0 & 0 & 1\\
        0 & 0 & 1 & 1\\
        0 & 1 & 2 & 1\\
        1 & 3 & 3 & 1\\
    \end{pmatrix}.
\end{align}
Note that the matrix $S$ in Eq.~\eqref{jordan-normal-j} is a lower triangular matrix with the structure of the Pascal's triangle or Pingala’s meru-prastara. Therefore, the Jordan normal form can be straightforwardly obtained from the transfer matrix $\mathbb{T}_{0}(\mu)$ by applying the mathematical structure of Pascal's triangle. In fact this statement is true for any order of EPs of the general transfer matrix given in Eq.~\eqref{general_transfer}, the matrix $S$ will always be a lower triangular matrix with the Pascal's triangle. Using the Jordan normal matrix $J$ given in Eq.~\eqref{jordan-normal-j}, one can compute,
\begin{align}
\label{Jpower}
    J^{-b}=e^{-i\pi b}\begin{pmatrix}
        1 &\,\,\,\,\,\, b & \frac{1}{2} b(b+1) & \frac{1}{6}b(b+1)(b+2)\\
        0 &\,\,\,\,\,\, 1 & b & \frac{1}{2} b(b+1) \\
        0 &\,\,\,\,\,\, 0 & 1 & b\\
        0 &\,\,\,\,\,\, 0 & 0 & 1 \\
    \end{pmatrix}.\,\,\,
\end{align}
Using the matrix $J^{-b}$ in Eq.~\eqref{Jpower} and $S$ in Eq.~\eqref{jordan-normal-j}, one can obtain the following required elements of $\langle i|\mathbb{T}_{0}(\mu)^{-b}|j\rangle$,
\begin{align}
    &\langle 3|\mathbb{T}_{0}(\mu)^{-b}|3\rangle =\frac{e^{-i\pi b}}{2}(2+b-2b^{2}-b^{3}),\label{X1}\\
    &\langle 3|\mathbb{T}_{0}(\mu)^{-b}|4\rangle =-\frac{e^{-i\pi b}}{6}\,b(b+1)(b+2),\label{X2}\\
    &\langle 4|\mathbb{T}_{0}(\mu)^{-b}|3\rangle =\frac{e^{-i\pi b}}{2}\, b(b+2)(b+3),\label{X3}\\
    &\langle 4|\mathbb{T}_{0}(\mu)^{-b}|4\rangle =\frac{e^{-i\pi b}}{6}\,(6+11b+6b^{2}+b^{3}),\label{X4}\\
    &\langle 3|\mathbb{T}_{0}(\mu)^{-b}|1\rangle=-\frac{e^{-i\pi b}}{6}\,b(b+1)(b+2)+\frac{e^{-i\pi b}}{2}\,b(b+1),\label{X5}\\
    &\langle 4|\mathbb{T}_{0}(\mu)^{-b}|1\rangle=\frac{e^{-i\pi b}}{6}\,b(b+1)(b+2).\label{X6}
\end{align}
From the expression of $A$ in Eq.~\eqref{A-matrix}, we finally obtain its elements as,
\begin{align}
    &A_{33}=\frac{1}{384}(-1)^{N}(N^{2}-1)\Big(i\gamma \,(N+3)^{2}(N+1)(N-3)-192\,(N+2)\Big), \label{S1}\\
    &A_{34}=\frac{1}{1152}(-1)^{N} (N+1)\Big(i\gamma \,(N+3)^{2}(N+1)(N-3)-192\,N(N+2)\Big),\label{S2}\\
    &A_{43}=-\frac{1}{384}(-1)^{N}(N+3)\Big(i\gamma \,(N+5)(N+3)(N+1)^{2}(N-3)-192\,N(N+2)\Big),\label{S3} \\
    &A_{44}=-\frac{1}{1152}(-1)^{N} (N+1)(N+3)\Big(i\gamma \,(N+5)(N+3)(N+1)^{2}(N-3)-192\,(N+2)\Big).\label{S4}
\end{align}
\end{widetext}
Using Eq.~\eqref{X1}-\eqref{X6} and Eq.~\eqref{S1}-\eqref{S4}, we obtain the system size scaling of the numerator of $|\langle 1 |M(\mu)^{-1}|N\rangle|$ in Eq.~\eqref{1MN} as 
$N^{3}/6$ for system size $N<B_4\,\mathcal{O}(1/\gamma^{1/3})$ and  as $\gamma\, N^{6}/(1152)$ for system size  $N>B_4\,\mathcal{O}(1/\gamma^{1/3})$ (where $B_4 \approx 6$). Similarly the denominator of $|\langle 1 |M(\mu)^{-1}|N\rangle|$ in Eq.~\eqref{1MN}  scales as $N^{4}/12$ for $N<B_4\,\mathcal{O}(1/\gamma^{1/3})$ and as
$\gamma N^{7}/1152$ for $N>B_4\,\mathcal{O}(1/\gamma^{1/3})$. Finally the scaling of $|\langle 1 |M(\mu)^{-1}|N\rangle|$ can be summarized as,
\begin{align}
\label{G-1N-per}
|g^{r}_{1N}(\mu)|=|\langle 1 |M(\mu)^{-1}|N\rangle| \sim
\begin{cases}
\frac{2}{N}\,\,\,\mathrm{for}\,\,N<B_4\,\mathcal{O}\Big(\frac{1}{\gamma^{1/3}}\Big) \\ \\
    \frac{1}{N}\,\,\,\mathrm{for}\,\,N>B_4\,\mathcal{O}\Big(\frac{1}{\gamma^{1/3}}\Big) 
\end{cases}
\end{align}
For the other matrix element $\langle 1 |M(\mu)^{-1}|m\rangle$ given in Eq.~\eqref{M1m}, the system size scaling can be similarly obtained as given as
\begin{align}
\label{G-1m-per}
   |g^{r}_{1m}(\mu)|= |\langle 1 |M(\mu)^{-1}|m\rangle| \sim
\begin{cases}
    \frac{N}{8}\,\,\,\mathrm{for}\,\,N<B_4\,\mathcal{O}\Big(\!\frac{1}{\gamma^{1/3}}\!\Big)\\ \\
    \frac{12}{\gamma N^{2}}\,\,\,\mathrm{for}\,\,N\!>\!B_4\,\mathcal{O}\Big(\!\frac{1}{\gamma^{1/3}}\!\Big) 
\end{cases}
\end{align}
Thus for $N<B_4\,\mathcal{O}(1/\gamma^{1/3})$, following Eqs.~\eqref{G-1N-per} and \eqref{G-1m-per} the conductance $G(\mu)$ given in Eq.~\eqref{conductance-within-band-app-gr} can be written as [correct upto $\mathcal{O}(\gamma_0)]$, (setting $\gamma_0=1)$
\begin{align}
    G(\mu)\sim \Big[\frac{4}{N^{2}}+\frac{\gamma}{128}N^{2}\Big].\label{cond-smallN}
\end{align}
From Eq.~\eqref{cond-smallN}, we find that when $N<A_4\,\mathcal{O}(1/\gamma^{1/4})$ where $A_4 \approx 5$, the conductance $G(\mu) \propto 1/N^{2}$ and for $A_4\,\mathcal{O}(1/\gamma^{1/4})<N<B_4\,\mathcal{O}(1/\gamma^{1/3})$, $G(\mu)\propto N^{2}$. Now for $N>(192/\gamma^{1/3})$, following Eqs.~\eqref{G-1N-per} and \eqref{G-1m-per}, the conductance $G(\mu)$ is obtained as [correct upto $\mathcal{O}(\gamma_0)]$ (setting $\gamma_0=1)$,
\begin{align}
    G(\mu) \sim \Big[\frac{1}{N^{2}}+\frac{72}{\gamma N^{4}}\Big].\label{cond-largeN}
\end{align}
From Eq.~\eqref{cond-largeN}, we find that when $B_4\,\mathcal{O}(1/\gamma^{1/3})<N<C_4\,\mathcal{O}(1/\gamma^{1/2})$ where $C_4 \approx 9$, the conductance $G(\mu) \propto 1/N^{4}$ and for $N>C_4\,\mathcal{O}(1/\gamma^{1/2})$, $G(\mu)\propto 1/N^{2}$. Thus, below we summarize our findings for $n=2$,
\begin{align}
\label{G-mu-n-2}
G(\mu) \propto
    \begin{cases}
        \frac{1}{N^{2}}\quad \mathrm{for}\,\, N<A_4\,\mathcal{O}\Big(\frac{1}{\gamma^{1/4}}\Big) \\ \\
        N^{2} \quad \mathrm{for}\,\, A_4\,\mathcal{O}\Big(\frac{1}{\gamma^{1/4}}\Big)<N<B_4\,\mathcal{O}\Big(\frac{1}{\gamma^{1/3}}\Big) \\ \\
        \frac{1}{N^{4}} \quad \mathrm{for}\,\, B_4\,\mathcal{O}\Big(\frac{1}{\gamma^{1/3}}\Big)<N<C_4\,\mathcal{O}\Big(\frac{1}{\gamma^{1/2} }\Big)\\ \\
        \frac{1}{N^{2}}\quad \mathrm{for}\,\, N>C_4\,\mathcal{O}\Big(\frac{1}{\gamma^{1/2}}\Big)
    \end{cases}
\end{align}
where one can estimate the constants $A_4 \approx 5 $, $B_4 \approx 6$, and $C_4 \approx 9$. This analytical result perfectly captures all the different transport regimes observed in Fig.~\ref{4th_order_local} for $n=2$.

\begin{figure}
    \centering
    \includegraphics[width=8.1cm]{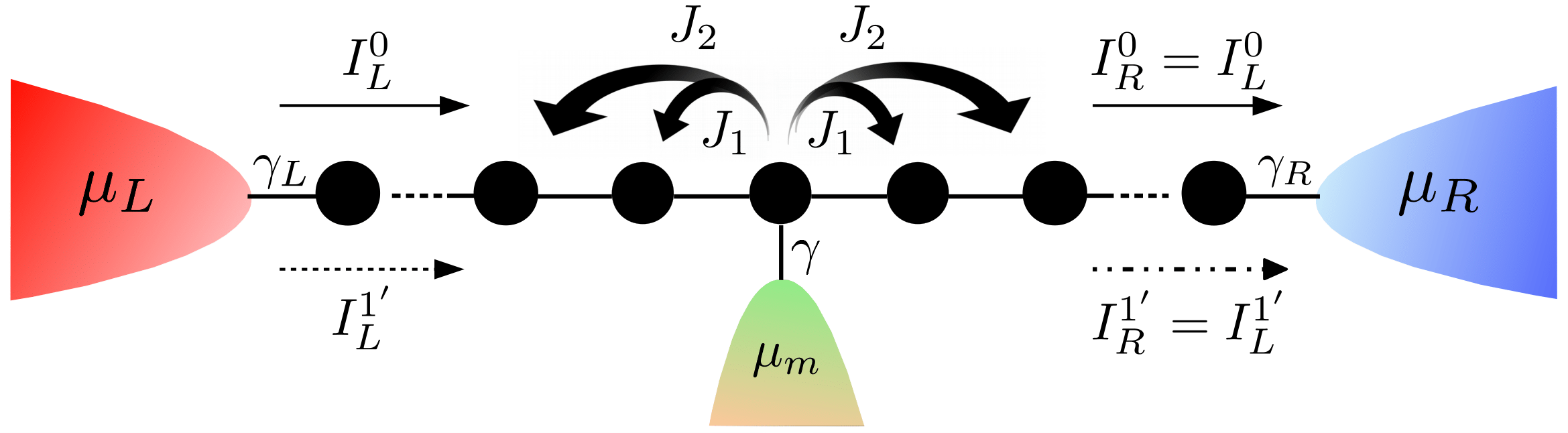}
    \caption{Single B\"uttiker probe connected at the middle site of the lattice with coupling strength $\gamma$. The chemical potential of the B$\ddot{\mathrm{u}}$ttiker probe is determined by demanding net zero particle current between the lattice and the reservoir and its expression is given by Eq.~\eqref{chem-pot-Buttiker}.}
    \label{Buttiker-probe}
\end{figure}

\section{Conductance in presence of single B\" uttiker probe at the middle site}
\label{appendix-B}
In this appendix, we show that under the condition $\gamma_L=\gamma_R$, the conductance expression in presence of a single B\" uttiker voltage probe \cite{Buttiker,PhysRevB.75.195110,korol2018probezt} located at the middle site of the lattice is exactly the same with that of a particle loss or a gain channel as given in Eq.~\eqref{conductance}.

We provide a schematic of a lattice connected to a B\" uttiker voltage probe  in Fig.~\ref{Buttiker-probe}. The current $I_p$ flowing out from the B$\ddot{\mathrm{u}}$ttiker probe to the system is zero at the steady state. The expression for this current is given as,
\begin{align}
        I_p=\int_{-\infty}^{+\infty} \frac{d\omega}{2\pi}\, &\big[\mathcal{P}_{L}(\omega)(f_p(\omega)-f_L (\omega))+ \nonumber \\
        &\mathcal{P}_{R}(\omega)(f_p(\omega)-f_R(\omega))\big]=0, \label{buttiker-probe-condition}
\end{align}
where $\mathcal{P}_{L}(\omega)$ and $\mathcal{P}_{R}(\omega)$ are the transmission function from the probe to the left and right reservoir, respectively. The subscript `$p$' in Eq.~\eqref{buttiker-probe-condition} refers to the `probe'.  Using the condition of zero current in Eq.~\eqref{buttiker-probe-condition} and assuming  zero temperature and linear response in chemical potential for the boundary reservoirs and the probe, we obtain the chemical potential at the middle site $m$ as,
\begin{align}
    \mu_m=\frac{\gamma_L\mu_L |\mathcal{G}^{r}_{1m}(\mu)|^{2}+\gamma_R\mu_R|\mathcal{G}^{r}_{mN}(\mu)|^{2}}{\gamma_L|\mathcal{G}^{r}_{1m}(\mu)|^{2}+\gamma_R|\mathcal{G}^{r}_{mN}(\mu)|^{2}}. \label{chem-pot-Buttiker}
\end{align}
As the probe is connected at the middle site $m$ and $\gamma_L=\gamma_R=\gamma_0$, we get $\mathcal{G}^{r}_{1m}(\mu)=\mathcal{G}^{r}_{mN}(\mu)$
[The detailed proof is given Sec.~\ref{sec-IV}, Eqs.~\eqref{B1}-\eqref{symmetry-Delta}].
As a result, the chemical potential at the middle site reduces to 
\begin{equation}
\label{mu-m1}
    \mu_m=\frac{\mu_L+\mu_R}{2}.
\end{equation}
In the presence of B\"uttiker probe, the steady state current flowing out of the left reservoir  is given as,
\begin{align}
    I_L=\gamma_0^{2}|\mathcal{G}^{r}_{1N}(\mu)|^{2}(\mu_L-\mu_R)+\gamma\gamma_0 |\mathcal{G}^{r}_{1m}(\mu)|^{2}(\mu_L-\mu_m) \label{Buttiker-current}
\end{align}
which is the same as the current $I_R$ in this case. Now substituting the expression for the chemical potential $\mu_m$ given in Eq.~\eqref{mu-m1}  in Eq.~\eqref{Buttiker-current}, we obtain the final expression for the conductance in presence of B$\ddot{\mathrm{u}}$ttiker probe as,
\begin{align}
    G=\frac{I_{L}}{\Delta\mu}=\gamma_0^{2}|\mathcal{G}^{r}_{1N}(\mu)|^{2}+\frac{\gamma\gamma_0}{2}|\mathcal{G}^{r}_{1m}(\mu)|^{2}, \label{conductance-Buttiker}
    \end{align}
where $\Delta \mu = \mu_L-\mu_R$. Therefore the conductance in presence of a single B\"uttiker voltage probe as given in Eq.~\eqref{conductance-Buttiker} is exactly the same expression as obtained in presence of local loss at the middle site [see Eq.~\eqref{conductance} in Sec.~\ref{sec-III}].

\section{Analytical proof of different transport regimes in presence of multiple non-extensive loss for range of hopping $n=2$}
\label{appendix-C}

In this appendix, we present the analytical proof that clearly demonstrates the emergence of different transport regimes at the band-edge of a finite range hopping model with range $n=2$ in presence of two lossy channels [see Fig.~\ref{multiple-nonextensive-loss}]. We recall that the schematic of this setup is given in the top panel of Fig.~\ref{schematic-extensive-loss}. The expression of conductance in such a scenario is,
\begin{align}
G(\mu)=\gamma_0^{2}|\mathcal{G}^{r}_{1N}(\mu)|^{2}+\frac{\gamma\gamma_0}{2}\sum_{l=\pm 1} |\mathcal{G}^{r}_{1 ,m-l}(\mu)|^{2},
\label{conductance-2loss}
\end{align}
where $l \pm 1$ denotes two lossy channels, placed symmetrically about the middle site. As done in appendix ~\ref{appendix-A}, to explain the system size scaling of the conductance $G(\mu)$, we again calculate $g^{r}(\mu)$ where we ignore boundary reservoir induced self energies $\Sigma_L$ and $\Sigma_R$. Thus, $g^{r}(\mu)$ is defined as,
\begin{align} g^{r}(\mu)=\frac{1}{J_2} \big[M(\mu)\big]^{-1}\,,\,\,\, M(\mu)=\frac{1}{J_2}\big[\mu\mathbb{I}_{N}-h_{S}+i\Gamma\big],
\end{align}
where the matrix elements of $\Gamma$ in presence of two lossy channels at the sites $m+1$ and $m-1$ are 
\begin{equation}
    \Gamma_{ij}=\frac{\gamma}{2}(\delta_{i,m+1}\delta_{j,m+1}+\delta_{i,m-1}\delta_{j,m-1}).
\end{equation}
The conductance expression in Eq.~\eqref{conduc-2loss} in terms of $g^{r}(\mu)$ reduces to,
\begin{align}
G(\mu)=\gamma_0^{2}|{g}^{r}_{1N}(\mu)|^{2}+\frac{\gamma\gamma_0}{2}\sum_{l=\pm 1} |{g}^{r}_{1,m-l}(\mu)|^{2},
\label{conductance-2loss-gr}
\end{align}
and is correct upto $\mathcal{O}(\gamma_0)$.

To compute the inverse of the matrix $M(\mu)$ \cite{LAVIS1997137}, we follow the similar prescription described in Appendix.~\ref{appendix-A} in Eqs.~\eqref{id-rel}-\eqref{VN-V0} and obtain,
\begin{align}
    &\mathbb{T}_{0}(\mu)\,V_{i}(j)=V_{i-1}(j)-\delta_{ij}\mathbb{I}_{4}|1\rangle \,\,\quad \mathrm{for}\,\, i\neq m\pm 1 \label{TM-rel-other-2loss}\\
    &\Tilde{\mathbb{T}}(\mu)\,V_{i}(j)=V_{i-1}(j)-\delta_{ij}\mathbb{I}_{4}|1\rangle \,\,\quad \mathrm{for}\,\, i= m\pm 1, \label{TM-rel-middle-2loss}
\end{align}
where recall that ${\mathbb{T}}_0(\mu)$ is the transfer matrix for $n=2$ as given in Eq.~\eqref{4th-order-TM1}. $\Tilde{\mathbb{T}}(\mu)=\Big[\mathbb{T}_{0}(\mu)-\frac{i\gamma}{2J_2}R\Big]$ with $R$ being a $4\times 4$ matrix with elements $R_{ij}=\delta_{1i}\delta_{2j}$. $V_{i}(j)$ is defined in Eq.~\eqref{V-vector}. Using Eqs.~\eqref{TM-rel-other-2loss}, \eqref{TM-rel-middle-2loss}, we obtain
\begin{widetext}
\begin{align}
    &V_{N}(N)=\big[\mathbb{T}_{0}(\mu)\big]^{-\frac{N-3}{2}} \,\Tilde{\mathbb{T}}(\mu)^{-1}\mathbb{T}_0(\mu)^{-1}\Tilde{\mathbb{T}}(\mu)^{-1}\big[\mathbb{T}_{0}(\mu)\big]^{-\frac{N-3}{2}} V_{0}(N) - \big[\mathbb{T}_{0}(\mu)\big]^{-1} |1\rangle, \label{TM-rel-other1-2loss} \\
    &V_{N}(m+1)=\big[\mathbb{T}_{0}(\mu)\big]^{-\frac{N-3}{2}} \,\Tilde{\mathbb{T}}(\mu)^{-1}\mathbb{T}_0(\mu)^{-1}\Tilde{\mathbb{T}}(\mu)^{-1}\big[\mathbb{T}_{0}(\mu)\big]^{-\frac{N-3}{2}} V_{0}(m+1) - \big[\mathbb{T}_{0}(\mu)\big]^{-\frac{N-3}{2}}\Tilde{\mathbb{T}}(\mu)^{-1} |1\rangle, \label{TM-rel-middle1-2loss}\\
    &V_{N}(m-1)=\big[\mathbb{T}_{0}(\mu)\big]^{-\frac{N-3}{2}} \,\Tilde{\mathbb{T}}(\mu)^{-1}\mathbb{T}_0(\mu)^{-1}\Tilde{\mathbb{T}}(\mu)^{-1}\big[\mathbb{T}_{0}(\mu)\big]^{-\frac{N-3}{2}} V_{0}(m+1) - \big[\mathbb{T}_{0}(\mu)\big]^{-\frac{N-3}{2}}\Tilde{\mathbb{T}}(\mu)^{-1}\mathbb{T}_{0}(\mu)^{-1} |1\rangle. \label{TM-rel-middle2-2loss}
\end{align}
We once again use the fact that $V_{N}(j)$ has zeros in its last two elements as shown in Eq.~\eqref{VN-V0}. Using Eqs.~\eqref{TM-rel-other1-2loss}-\eqref{TM-rel-middle2-2loss}, we write
\begin{align}
    &\sum_{l=1,2} \langle s+l | B | 2+l\rangle \langle l|M(\mu)^{-1}|N\rangle - \langle s+l| \mathbb{T}_{0}(\mu)^{-1}|1\rangle =0,\,\,\quad \mathrm{where}\,\, s=1,2\label{sum1-1}\\
    &\sum_{l=1,2} \langle s+l | B | 2+l\rangle \langle l|M(\mu)^{-1}|m+1\rangle - \langle s+l| \mathbb{T}_{0}(\mu)^{-\frac{N-3}{2}}\Tilde{\mathbb{T}}(\mu)^{-1} |1\rangle =0,\,\,\quad \mathrm{where}\,\, s=1,2\label{sum1-2} \\
    &\sum_{l=1,2} \langle s+l | B | 2+l\rangle \langle l|M(\mu)^{-1}|m-1\rangle - \langle s+l| \mathbb{T}_{0}(\mu)^{-\frac{N-3}{2}}\Tilde{\mathbb{T}}(\mu)^{-1}\mathbb{T}_{0}(\mu)^{-1} |1\rangle =0.\,\,\quad \mathrm{where}\,\, s=1,2.\label{sum1-3} 
\end{align}
Here the matrix $B$ is given as,
\begin{align}
\label{B-mat}
B=\big[\mathbb{T}_{0}(\mu)\big]^{-\frac{N-3}{2}} \,\Tilde{\mathbb{T}}(\mu)^{-1}\mathbb{T}_0(\mu)^{-1}\Tilde{\mathbb{T}}(\mu)^{-1}\big[\mathbb{T}_{0}(\mu)\big]^{-\frac{N-3}{2}}.
\end{align} 
Henceforth we follow a similar procedure as done in Appendix \ref{appendix-A} in Eq.~\eqref{procedure-A}. Using Eqs.~\eqref{sum1-1}-\eqref{sum1-3} we write,
\begin{equation}
    \langle 1| M(\mu)^{-1}|N\rangle=\frac{B_{44} \langle 3|\mathbb{T}_{0}(\mu)^{-1}|1\rangle\,-\,B_{34}\langle 4|\mathbb{T}_{0}(\mu)^{-1}|1\rangle}{B_{33}B_{44}-B_{34}B_{43}}, \label{1MN-2loss}
\end{equation}
\begin{equation}
    \langle 1| M(\mu)^{-1}|m+1\rangle=\frac{B_{44} \langle 3|\mathbb{T}_{0}(\mu)^{-\frac{N-3}{2}}\Tilde{\mathbb{T}}(\mu)^{-1} |1\rangle\,-\,B_{34}\langle 4|\mathbb{T}_{0}(\mu)^{-\frac{N-3}{2}}\Tilde{\mathbb{T}}(\mu)^{-1} |1\rangle}{B_{33}B_{44}-B_{34}B_{43}},\label{1m1-2loss}
\end{equation}
\begin{equation}
    \langle 1| M(\mu)^{-1}|m-1\rangle=\frac{B_{44} \langle 3|\mathbb{T}_{0}(\mu)^{-\frac{N-3}{2}}\Tilde{\mathbb{T}}(\mu)^{-1}\mathbb{T}_0(\mu)^{-1}|1\rangle\,-\,B_{34}\langle 4|\mathbb{T}_{0}(\mu)^{-\frac{N-3}{2}}\Tilde{\mathbb{T}}(\mu)^{-1}\mathbb{T}_0(\mu)^{-1} |1\rangle}{B_{33}B_{44}-B_{34}B_{43}}.
    \label{1m2-2loss}
\end{equation}
The required elements of the matrix $B$ given in Eq.~\eqref{B-mat} is obtained using the Jordan normal form of the transfer matrix $\mathbb{T}_{0}(\mu)$ given in Eq.~\eqref{jordan-normal-j} [recall that we set $\mu=3 J_1/2$ corresponding to  the upper band-edge energy and $J_1=1$ here]. We obtain 
\begin{align}
    &B_{33}=\frac{1}{192}(-1)^{N}(N^{2}-1)\Big[i\gamma(N+3)(N+5)(N^{2}-4N-1)+\gamma^{2}(N^{2}-1)(N-3)(N-5)-96(N+2)\Big], \label{b33} \\
    &B_{34}=\frac{1}{576}(-1)^{N}(N+1)\Big[i\gamma(N^{2}-9)(N^{2}-1)(N+5)+\gamma^{2}(N+1)(N^{2}-4N+3)^{2}-96N(N+2)\Big], \label{b34} \\
    &B_{43}=-\frac{1}{192}(-1)^{N}(N+3)\Big[i\gamma(N^{2}-1)(N+5)(N^{2}+2N-19)+\gamma^{2}(N-5)(N^{2}-1)^{2}-96N(N+2)\Big], \label{b43} \\
    &B_{44}=-\frac{1}{576}(-1)^{N}(N+1)(N+3)\Big[i\gamma (N+5)(N-1)(N^{2}+4N-9)+\gamma^{2}(N-1)(N-3)(N^{2}-1)-96N(N+2)\Big].\label{b44}
\end{align}
The other two important quantities that we need to compute to obtain the scaling of $\langle 1|M(\mu)^{-1}|m \pm 1\rangle$ following Eqs.~\eqref{1m1-2loss} and \eqref{1m2-2loss} are,
\begin{align}
    &\langle 3| \mathbb{T}_{0}(\mu)^{-\frac{N-3}{2}}\Tilde{\mathbb{T}}(\mu)^{-1} |1\rangle=\frac{1}{48}(-1)^{\frac{N+1}{2}}(N^{2}-1)(N-3),\label{TFULL1} \\
    &\langle 4| \mathbb{T}_{0}(\mu)^{-\frac{N-3}{2}}\Tilde{\mathbb{T}}(\mu)^{-1} |1\rangle=-\frac{1}{48}(-1)^{\frac{N+1}{2}}(N^{2}-1)(N+3),\label{TFULL2} \\
    &\langle 3| \mathbb{T}_{0}(\mu)^{-\frac{N-3}{2}}\Tilde{\mathbb{T}}(\mu)^{-1} \mathbb{T}_{0}^{-1}|1\rangle=-\frac{1}{48}(-1)^{\frac{N+1}{2}}(N^{2}-1)(N+3),\label{TFULL3} \\
    &\langle 4| \mathbb{T}_{0}(\mu)^{-\frac{N-3}{2}}\Tilde{\mathbb{T}}(\mu)^{-1} \mathbb{T}_{0}^{-1}|1\rangle=\frac{1}{48}(-1)^{\frac{N+1}{2}}(N+1)(N+3)(N+5).
    \label{TFULL4} 
\end{align}
\end{widetext}
Using Eqs.~\eqref{b33}-\eqref{b44}, we obtain the scaling of $\langle 1|M(\mu)^{-1}|N\rangle$, $\langle 1|M(\mu)^{-1}|m+1\rangle$ and $\langle 1|M(\mu)^{-1}|m-1\rangle$ as (henceforth we set $\gamma_0=1$), 
\begin{align}
\label{1MN-2loss}
|g_{1N}^{r}(\mu)| \sim
    \begin{cases}
       \frac{2}{N} \,\,\,\,\,\,\,\,\,\,\,\mathrm{for}\,\, N<B'_{4}\,\mathcal{O}\Big(\frac{1}{\gamma^{1/3}}\Big)\\
       \frac{24}{\gamma N^4}\,\,\,\,\,\,\mathrm{for}\,\, B'_4\,\mathcal{O}\Big(\frac{1}{\gamma^{1/3}}\Big)<N<\Tilde{B}_4\,\mathcal{O}\Big(\frac{1}{\gamma^{1/3}}\Big)\\
       \frac{1}{N} \,\,\,\,\,\,\,\,\,\,\,\mathrm{for}\,\, \Tilde{A}_4\,\mathcal{O}\Big(\frac{1}{\gamma^{1/3}}\Big)<N<\Tilde{D}_{4}\,\mathcal{O}\Big(\frac{1}{\gamma}\Big)\\
       \frac{4}{N^{2}} \,\,\,\,\,\,\,\,\,\mathrm{for}\,\, N>\Tilde{D}_{4}\,\mathcal{O}\Big(\frac{1}{\gamma}\Big)
    \end{cases}
\end{align}
where $B'_4 \approx 4$, $\Tilde{B}_4 \approx 5$, and $\Tilde{D}_4 \approx 4$.
\begin{align}
\label{1mm1-2loss}
|g_{1,m+1}^{r}(\mu)| \sim
    \begin{cases}
       \frac{N}{8} \,\,\,\,\,\,\,\,\,\,\mathrm{for}\,\, N<B'_4\,\mathcal{O}\Big(\frac{1}{\gamma^{1/3}}\Big)\\
       \frac{6}{\gamma N^{2}} \,\,\,\,\,\mathrm{for}\,\, B'_4\,\mathcal{O}\Big(\frac{1}{\gamma^{1/3}}\Big)<N<\Tilde{C}_4\,\mathcal{O}\Big(\!\frac{1}{\gamma^{1/2}}\!\Big)\\
       \frac{1}{4}N^0 \,\,\,\,\mathrm{for}\,\, \Tilde{C}_4\,\mathcal{O}\Big(\frac{1}{\gamma^{1/2}}\Big)<N<\Tilde{D}_4\,\mathcal{O}\Big(\frac{1}{\gamma}\Big)\\
       \frac{1}{\gamma N} \,\,\,\,\,\,\,\,\mathrm{for}\,\, N>\Tilde{D}_4\,\mathcal{O}\Big(\frac{1}{\gamma}\Big)
    \end{cases}
\end{align}
where $\Tilde{C}_4 \approx 5$.
\begin{align}
\label{1mm2-2loss}
|g_{1,m-1}^{r}(\mu)| \sim
    \begin{cases}
       \frac{N}{8} \,\,\,\,\,\,\,\,\,\,\,\mathrm{for}\,\, N<B'_4\,\mathcal{O}\Big(\frac{1}{\gamma^{1/3}}\Big)\\
       \frac{6}{\gamma N^{2}} \,\,\,\,\,\,\mathrm{for}\,\, B'_4\,\mathcal{O}\Big(\frac{1}{\gamma^{1/3}}\Big)<N<\Tilde{D}_4\,\mathcal{O}\Big(\frac{1}{\gamma}\Big)\\
       \frac{24}{\gamma^{2}N^{3}}\,\,\,\,\mathrm{for}\,\, \Tilde{D}_4\,\mathcal{O}\Big(\frac{1}{\gamma}\Big)<N<D'_4\,\Big(\frac{1}{\gamma}\Big)\\
       \frac{4}{\gamma N} \,\,\,\,\,\,\,\,\,\mathrm{for}\,\, N>D'_4\,\mathcal{O}\Big(\frac{1}{\gamma}\Big)
    \end{cases}
\end{align}
Now using Eqs.~\eqref{1MN-2loss}-\eqref{1mm2-2loss}, we obtain the different transport regimes of conductance $G(\mu)$. First let us consider the window $N<B'_4\,(1/\gamma^{1/3})$, the conductance $G(\mu)$ is,
\begin{align}
    G(\mu)\sim \Big[\frac{4}{N^{2}}+\frac{\gamma N^{2}}{64}\Big]. \label{conduc-2loss}
\end{align}
From Eq.~\eqref{conduc-2loss}, we find that when $N<A'_4\,\mathcal{O}(1/\gamma^{1/4})$ where $A'_4 \approx 4$, the conductance $G(\mu)\propto \frac{1}{N^{2}}$  and for $A'_4\,\mathcal{O}(1/\gamma^{1/4})<N<B'_4\,\mathcal{O}(1/\gamma^{1/3})$, $G(\mu)\propto N^{2}$. Next for the window, $B'_{4}\,\mathcal{O}(1/\gamma^{1/3})<N<\Tilde{B}_4\,\mathcal{O}(1/\gamma^{1/3})$ (recall that $B'_4 \approx 4$ and $\Tilde{B}_4 \approx 5$), the conductance $G(\mu)$ scales as,
\begin{align}
    G(\mu) \sim \Big[\frac{576}{\gamma^{2}N^8}+\frac{36}{\gamma N^4}\Big] \propto \frac{1}{N^{4}}.
\end{align}
Next, for the range of $N$ i.e., $\Tilde{B}_4\,\mathcal{O}(1/\gamma^{1/3})<N<\Tilde{C}_4\,\mathcal{O}(1/\gamma^{1/2})$ (recall that $\Tilde{C}_4 \approx 5$), $G(\mu)$ is given as,
\begin{align}
    G(\mu)\sim \Big[\frac{1}{N^{2}}+\frac{36}{\gamma N^{4}}\Big] \propto \frac{1}{N^{4}}.
\end{align}
Now for the range of $\Tilde{C}_4\,\mathcal{O}(1/\gamma)^{1/2}<N<D'_4\,\mathcal{O}(1/\gamma)$, the conductance is,
\begin{align}
    G(\mu)\sim \Big[\frac{1}{N^{2}}+\frac{\gamma}{32}N^0+\frac{18}{\gamma N^{4}}\Big].
\end{align}
For $\Tilde{C}_4\,\mathcal{O}(1/\gamma^{1/2})<N<C'_4\,\mathcal{O}(1/\gamma^{1/2})$, the conductance $G(\mu)\propto 1/N^{2}$ but this window is very small and difficult to observe and hence we ignore this regime.  Next for $C'_4\,\mathcal{O}(1/\gamma^{1/2})<N<D'_4\,\mathcal{O}(1/\gamma)$, the conductance scales as $N^{0}$. After this ballistic $N^0$ regime, for the window $N>D'_4\,\mathcal{O}(1/\gamma)$, the conductance $G(\mu)$ scales as, 
\begin{align}
G(\mu)\propto \frac{1}{N^{2}}.
\end{align}
Thus, we summarize the result for $n=2$ in presence of two lossy channels,
\begin{align}
G(\mu)\propto 
    \begin{cases}
        \frac{1}{N^{2}} \,\,\,\,\mathrm{for}\,\,N<A'_4\,\mathcal{O}\Big(\frac{1}{\gamma^{1/4}}\Big)\\
        N^{2} \,\,\,\,\mathrm{for}\,\,A'_4\,\mathcal{O}\Big(\frac{1}{\gamma^{1/4}}\Big)<N<B'_4\,\mathcal{O}\Big(\frac{1}{\gamma^{1/3}}\Big)\\
        \frac{1}{N^{4}}\,\,\,\,\mathrm{for}\,\,B'_4\,\mathcal{O}\Big(\frac{1}{\gamma^{1/3}}\Big)<N<C'_4\,\mathcal{O}\Big(\frac{1}{\gamma^{1/2}}\Big)\\
        N^{0}\,\,\,\,\mathrm{for}\,\,\,\,\,C'_4\,\mathcal{O}\Big(\frac{1}{\gamma^{1/2}}\Big)<N<D'_4\,\mathcal{O}\Big(\frac{1}{\gamma}\Big)\\
        \frac{1}{N^{2}}\,\,\,\,\mathrm{for}\,\, N>D'_4\,\mathcal{O}\Big(\frac{1}{\gamma}\Big)
    \end{cases}
\end{align}
where $A'_4 \approx 4$, $B'_4 \approx 4$, $C'_4 \approx 7$ and $D'_4 \approx 6$. This analytical result perfectly captures all the different transport regimes observed in Fig.~\ref{multiple-nonextensive-loss} for $n=2$.

\bibliography{references}
\end{document}